\let\includefigures=\iffalse
%
\let\useblackboard=\iftrue
%
%
\newfam\black
\input harvmac
\useblackboard
\message{If you do not have msbm (blackboard bold) fonts,}
\message{change the option at the top of the tex file.}
\font\blackboard=msbm10
\font\blackboards=msbm7
\font\blackboardss=msbm5
\textfont\black=\blackboard
\scriptfont\black=\blackboards
\scriptscriptfont\black=\blackboardss
\def\Bbb#1{{\fam\black\relax#1}}
\else
\def\Bbb{\bf}
\fi
\def\Cb{{\Bbb C}}
\def\Rb{{\Bbb R}}
\def\Zb{{\Bbb Z}}
\def\Tb{{\Bbb T}}
\def\Ac{{\cal A}}
\def\Gc{{\cal G}}
\def\Hc{{\cal H}}
\def\Mc{{\cal M}}

\def\Vc{{\cal V}}
\def\nl{\hfill\break}
\def\p{\partial}
\def\I{I}

\def\II{\relax{I\kern-.10em I}}
\def\IIa{{\II}a}
\def\IIb{{\II}b}
\def\sgn{{\rm sgn\ }}
\def\Tr{{\rm Tr\ }}
\def\p{{\partial}}
\def\pb{{\bar\partial}}
\def\build#1_#2^#3{\mathrel{
\mathop{\kern 0pt#1}\limits_{#2}^{#3}}}
\def\boxit#1#2{\setbox1=\hbox{\kern#1{#2}\kern#1}%
\dimen1=\ht1 \advance\dimen1 by #1 \dimen2=\dp1 \advance\dimen2 by #1
\setbox1=\hbox{\vrule height\dimen1 depth\dimen2\box1\vrule}%
\setbox1=\vbox{\hrule\box1\hrule}%
\advance\dimen1 by .4pt \ht1=\dimen1
\advance\dimen2 by .4pt \dp1=\dimen2 \box1\relax}
\Title{\vbox{\baselineskip12pt\hbox{hep-th/9711162}
\hbox{IHES/P/97/82}
\hbox{RU-97-94}}}
{\vbox{
\centerline{Noncommutative Geometry and Matrix Theory:}
\medskip
\centerline{Compactification on Tori} }}
\centerline{Alain Connes$^1$, Michael R. Douglas$^{1,2}$
 and Albert Schwarz$^{1,2,3}$}
\medskip\centerline{$^1$ Institut des Hautes \'Etudes Scientifiques}
\centerline{Le Bois-Marie, Bures-sur-Yvette, 91440 France}
\medskip
\centerline{$^2$ Department of Physics and Astronomy}
\centerline{Rutgers University }
\centerline{Piscataway, NJ 08855--0849 USA}
\medskip
\centerline{$^3$ Department of Mathematics}
\centerline{University of California}
\centerline{Davis, CA 95616 USA}
\medskip
\centerline{\tt connes@ihes.fr, douglas@ihes.fr, asschwarz@ucdavis.edu}
\bigskip
We study toroidal compactification of Matrix theory,
using ideas and results of non-commutative geometry.
We generalize this to compactification on
the noncommutative torus, explain the classification of these backgrounds,
and argue that they correspond in supergravity
to tori with constant background three-form tensor field.
The paper includes an introduction for mathematicians
to the IKKT formulation of Matrix theory
and its relation to the BFSS Matrix theory.

\Date{November 1997}
%
\lref\bfss{T. Banks, W. Fischler, S. H. Shenker and L. Susskind,
Phys. Rev. D55 (1997) 5112-5128; hep-th/9610043.}
\lref\dhn{B. de Wit, J. Hoppe and H. Nicolai,
Nucl.Phys. {\bf B 305 [FS 23]} (1988) 545.}
\lref\town{P. Townsend, to appear in the Strings '97 proceedings.}
\lref\bst{E. Bergshoeff, E. Sezgin and P. K. Townsend,
Phys. Lett. 189B (1987) 75;
Ann. Phys. 185 (1988) 330.}
\lref\DLN{B. de Wit, M. L\"uscher and H. Nicolai,
Nucl.Phys. {\bf B 305 [FS 23]} (1988) 545.}
\lref\HT{C. Hull and P. K. Townsend}
\lref\Witone{E. Witten}
\lref\DLP{J.~Dai, R.~G.~Leigh and J.~Polchinski,
Mod. Phys. Lett. {\bf A4} (1989) 2073.}
\lref\Pol{J.~Polchinski, Phys.~Rev.~Lett.~{\bf 75} (1995) 4724-4727;
hep-th/9510017.}
\lref\susslc{L. Susskind, hep-th/9704080.}
\lref\dkps{M. R. Douglas, D. Kabat, P. Pouliot and S. Shenker,
Nucl. Phys. B485 (1997) 85-127; hep-th/9608024.}
\lref\doug{M. R. Douglas, to appear in the '97 Les Houches proceedings}
\lref\ikkt{N. Ishibashi, H. Kawai, Y. Kitazawa, A. Tsuchiya,
Nucl.Phys. B498 (1997) 467-491.}
\lref\grt{O. Ganor, S. Ramgoolam and W. Taylor,
Nucl. Phys. B492 (1997) 191-204; hep-th/9611202.}
\lref\suss{L. Susskind, hep-th/9611164.}
\lref\Banks{T. Banks, ``Matrix Theory,'' hep-th/9710231.}
\lref\connes{A. Connes, {\it Noncommutative Geometry,} Academic Press, 1994.}
\lref\connesym{A. Connes,
Comm. Math. Phys. 182 (1996) 155-176; hep-th/9603053.}
\lref\connesmath{A. Connes, C. R. Acad. Sci. Paris S\'er. A-B 290 (1980)
A599-A604; \nl
M. Pimsner and D. Voiculescu, J. Operator Theory 4 (1980) 93--118; \nl
A. Connes and M. Rieffel, ``Yang-Mills for noncommutative two-tori,''
in Operator Algebras and Mathematical Physics (Iowa City, Iowa, 1985),
pp. 237--266, Contemp. Math. Oper. Algebra. Math. Phys. 62,
AMS 1987; \nl
M. Rieffel, ``Projective modules over higher-dimensional
non-commutative tori,'' Can. J. Math, 40 (1988) 257--338.}
\lref\taylor{W. Taylor, Phys. Lett. B394 (1997) 283-287; hep-th/9611042.}
\lref\defquant{M. De Wilde and P. Lecomte, Lett. Math. Phys. 7 (1983) 487; \nl
B. V. Fedosov, J. Diff. Geom. 40 (1994) 213; \nl
M. Kontsevich, q-alg/9709040.}
\lref\dhn{B. de Wit, J. Hoppe and H. Nicolai,
Nucl. Phys. B305 [FS 23] (1988) 545.}
\lref\ncfields{C. Klim{\u c}ik, hep-th/9710153.}
\lref\hoppe{J. Hoppe, Phys. Lett. B250 (1990) 44.}
\lref\zfc{C. Zachos, D. Fairlie and T. Curtright, hep-th/9709042.}
\lref\baake{M. Baake, P. Reinicke and V. Rittenberg, J. Math Phys. 26 (1985)
1070; \nl
R. Flume, Ann. Phys. 164 (1985) 189; \nl
M. Claudson and M. B. Halpern; Nucl. Phys. B250 (1985) 689.}
\lref\tfour{
M. Rozali, Phys. Lett. B400 (1997) 260-264, hep-th/9702136; \nl
M. Berkooz, M. Rozali and N. Seiberg, hep-th/9704089.}
\lref\hack{F. Hacquebord and H. Verlinde, hep-th/9707179.}
\lref\townsend{P. K. Townsend, hep-th/9612121;
to appear in procs. of the ICTP June 1996 summer school.}
\lref\moore{G. Moore, ``Finite in All Directions,'' hep-th/9305139.}
\lref\schwarz{A. Schwarz, in preparation. }
\lref\nbi{A. Tseytlin, hep-th/9701125.}
\lref\witten{E. Witten, Phys. Lett. 86B (1979) 283.}
\lref\freed{D. Freed, ``Locality and integration in topological field
theory,'' in Group Theoretical methods in Physics, Volume 2, eds. M.A. del
Olmo,
M. Santander and J. M. Guilarte, Ciemat, 1993, 35--54.}
\lref\caicedo{M. I. Caicedo, I. Martin and A. Restuccia, hep-th/9711122.}
\lref\hull{C. Hull, private communication.}
\lref\page{D. Page, Phys. Rev. D28 (1983) 2976.}
\newsec{Introduction}

The recent development of superstring theory has shown that these
theories are the perturbative expansions of a more general theory where
strings are on equal footing with their multidimensional
analogs (branes).
This theory is called M theory, where M
stands for ``mysterious'' or ``membrane.''

It was conjectured in \bfss\ that M theory can be defined as a
matrix quantum mechanics,
obtained from ten-dimensional supersymmetric
Yang-Mills (SYM) theory by means of reduction to
$0+1$ dimensional theory,
where the size of the matrix tends to infinity.
Another matrix model was
suggested in \ikkt; it can be obtained by reduction of
10-dimensional SYM theory to a point.
The two models, known as the BFSS Matrix model and the IKKT Matrix
model, are closely related.

The goal of the present paper is to formulate the IKKT and BFSS
Matrix models, to make more precise the relation
between these models, and to study their toroidal
compactifications.
We will describe a new kind of toroidal
compactification and show how the methods of noncommutative
geometry can be used to analyze them.  The paper is
self-contained and, we
hope, accessible both to physicists and mathematicians. A
mathematician can use it as a very short introduction to
Matrix theory.

In section 2 we describe the IKKT model and review the
relation of this model to Green-Schwarz superstring theory following \ikkt.

In section 3 we discuss toroidal compactification along the lines
of \refs{\taylor,\bfss,\grt} (see \Banks\ for other references).
Compactifying one dimension in the IKKT model
leads to a formal relation to the BFSS model (known in the physics
literature as ``Eguchi-Kawai reduction'').
In two and more dimensions,
although we start with the same defining relations as \refs{\taylor,\bfss},
we show that
they admit more general solutions than previously considered.
These solutions exactly correspond to generalizing vector bundles
over the torus to those over the noncommutative torus.

This motivates the introduction of noncommutative geometry, and
we discuss the relevant ideas in section 4.
Quite strikingly, the defining relations of toroidal compactification
in the framework of \bfss\ are precisely the definition of
a connection on the noncommutative torus.
We describe two commutative tori naturally associated
to a noncommutative torus, one to its odd and one to its even
cohomology, leading to two commuting $SL(2,\Zb)$ actions on the
Teichmuller space.
The moduli space of constant curvature connections, associated to
the odd cohomology, will play the role of space-time, just as for
conventional toroidal compactification, while
the torus associated to the even cohomology and its associated
$SL(2,\Zb)$ has no direct analog in the commuting case.

In section 5 we discuss the new theories as gauge theories, using
an explicit Lagrangian written in conventional physical terms.
The Lagrangian is of the same general type
used to describe supermembranes in \dhn, with Poisson brackets
replaced by Moyal brackets.

In section 6 we propose a physical interpretation for the new
compactifications, in the context of the BFSS model as generalized
by Susskind \susslc.
The matrix theory hypothesis requires them
to correspond to solutions of eleven-dimensional supergravity,
with space-time determined as the moduli space of supersymmetric vacua.
This is the
moduli space of constant curvature connections and thus they
must be toroidal compactifications but with non-zero background fields
consistent with supersymmetry.

We argue that deforming the commutative torus to the noncommutative torus
corresponds to turning on a constant
background three-form potential $C_{ij-}$.
In the case of the noncommutative two-torus,
we argue that the additional $SL(2,\Zb)$ duality symmetry predicted
by the noncommutative geometry approach is present, and corresponds to
T-duality on a two-torus including the compact null dimension.  We
check that the BPS mass formula as well as the string world-sheet
description respect this symmetry.

Section 7 contains conclusions.

\newsec{Introduction to the IKKT model}

Our starting point will be the IKKT model in its Euclidean version.
We define this model by giving a complex supermanifold as configuration
space and an action functional $S$, considered as a
holomorphic function on this space.
All physical quantities can be
expressed as integrals with an integrand containing $\exp (-S)$.

We first make a technical remark which will
permit us to avoid complications
related to the absence of Majorana-Weyl spinors in the Euclidean
setting.
As usual to integrate a holomorphic function we
should specify a real cycle (real slice), but the integral
depends only on the homology class of the cycle (in
non-compact case one should consider an appropriate relative
homology). It is important to emphasize that the integral of
a holomorphic function over a complex supermanifold equipped
with a complex volume element does not depend on the choice
of ``odd part'' of a real slice (i.e. to define the integral
we should specify only the real slice in the body of
supermanifold). We work with Weyl spinors (i.e. with quantities
that transform according to one of the irreducible
representations in the decomposition of the spinor representation
of $SO(10,\Cb)$ into left and right parts). Due to the absence
of Majorana-Weyl spinors there is no
$SO(10,\Cb)$-invariant real slice, but this is irrelevant for us.

The symbol $\Pi$ is used to denote the parity reversion (e.g.
$\Pi R^{m\mid n} = R^{n \mid m})$.

Let us consider an action functional
\eqn\action{
S = R \sum_{i,j} \langle [X_i ,X_j],[X_i ,X_j]
\rangle + 2R \sum_i \langle \Psi^{\alpha} ,
\Gamma_{\alpha \beta}^i [X_i , \Psi^{\beta}] \rangle
}
where $X_i$, $i=0,1,\ldots ,9$ are elements of a complex Lie
algebra $\Gc$ equipped with an invariant bilinear inner product
$\langle , \rangle$, $\Psi^{\alpha}$, $\alpha = 1, \ldots
,16$ are elements of $\Pi\Gc$ and $\Gamma_{\alpha \beta}^i$
are ten-dimensional Dirac matrices.
$R$ is a constant of
normalization whose significance will be explained below.

The functional \action\ is
invariant with respect to the action of complex orthogonal
group $SO(10,\Cb)$ if $X_0 , \ldots ,X_9$ transform as a vector
and $\Psi^1 , \ldots ,\Psi^{16}$ as a Weyl spinor.
More
precisely if $\Cb^{10}$ stands for the space of fundamental
representation of $SO (10,\Cb)$ and $S$ for the space of
irreducible sixteen-dimensional two-valued representation of
$SO (10,\Cb)$, then $X \in\Gc \otimes\Cb^{10}$ and $\Psi \in \Pi
\Gc \otimes S$. Elements $X_0 , \ldots ,X_9$ and $\Psi^1 ,
\ldots , \Psi^{16}$ are components of $X$ and $\Psi$ in
fixed bases in $\Cb^{10}$ and $S$ respectively.
Matrices $\Gamma_{\alpha \beta}^i$ correspond to the
operators $\Gamma^i$ acting in $S \oplus S^*$ and obeying $\Gamma^i
\Gamma^j + \Gamma^j \Gamma^i = 2 \delta^{ij}$.
(The operators $\Gamma^i$ act on the space of spinor representation
of orthogonal group. Taking into account that the number 10 has the
form $4n+2$ we see that this space can be decomposed into
direct sum of irreducible representation $S$ and dual representation
$S^*$. The operator $\Gamma_i$  acts from $S$ into $S^*$.
There is an invariant bilinear pairing between $S$ and $S^*$
which we have implicitly used in this formula.)

The functional \action\ is also invariant with respect to
infinitesimal gauge transformations $X_i \rightarrow
[U,X_i]$, $\Psi^{\alpha} \rightarrow [U, \Psi^{\alpha}]$
with $U \in \Gc$, and with respect to supersymmetry
transformations
\eqn\susy{\eqalign{
\delta^{(1)} X^i = & \ \varepsilon^\alpha \Gamma_{\alpha \beta}^i \Psi^\beta
\cr
\delta^{(1)} \Psi = & \ {1 \over 2} \, [X_i , X_j] \Gamma^{ij}
\varepsilon \cr
\delta^{(2)} X_i = & \ 0 \cr
\delta^{(2)} \Psi = & \ \xi \cdot \gamma \cr
}}
where $\varepsilon$ and $\xi$ are Weyl spinors (i.e.
$\varepsilon \in S$, $\xi \in S$) and $\gamma$ belongs to the
center of ${\Gc}$. Here $\Gamma^{ij} = [\Gamma^i ,
\Gamma^j]$, $\Gamma_i = \Gamma^i$. (We fixed an orthonormal
basis in $\Cb^{10}$ and therefore the distinction between
upper and lower indices is irrelevant.)

If ${\Ac}$ is an associative algebra with trace then the
corresponding Lie algebra (i.e. ${\cal A}$ equipped with the
operation $[x,y] = xy - yx$) has an invariant inner product
$\langle x,y \rangle = \Tr xy$. In particular, we can
consider the algebra of complex $N \times N$ matrices:
${\Ac} = {\rm Mat}_N (\Cb)$. Then \action\ is
the action functional of the IKKT Matrix model suggested in
\ikkt.

Another term invariant under the symmetry \susy\ which can be
added to the action in this case is
\eqn\topterm{
S_2 = \sum_{i,j} \gamma_{ij} \Tr [X^i,X^j].
}
Although it vanishes for finite $N$, it will play a role
in the limit $N \rightarrow \infty$.

The functional \action\ is a holomorphic
function on the superspace $\Cb^{10|16}
\times {\rm Mat}_N$ (superspace of states);
i.e. on the space of rows $(X_0 , \ldots ,X_9 , \Psi^1
, \ldots , \Psi^{16})$ where $X_i$ are even complex $N
\times N$ matrices and $\Psi^{\alpha}$ are odd complex $N
\times N$ matrices.
Physical quantities (e.g., correlation functions)
are defined in terms of an integral over
a real slice in the body
of this superspace; for example we can require hermiticity of the
matrices $X_0 , \ldots ,X_9$.

\subsec{Physical interpretation}

It was conjectured in \ikkt\  that this functional integral
in the limit $N,R \rightarrow \infty$ with $N/R$ fixed
can be used as a non-perturbative definition
of the type \IIb\ superstring theory.
This conjecture is prompted by the remark that the
action functional \action\ is closely related to the
action functional of Green-Schwarz superstring in the case
when ${\Gc} = C^{\infty} (M)$ is a Lie algebra of complex
smooth functions on two-dimensional compact manifold $M$
equipped with a symplectic structure. (The commutator is
given by the Poisson bracket, the inner product $(f,g)$ is
defined as an integral of $f \cdot g$ over $M$.) More
precisely, one should consider the set ${\Vc}$ of all
symplectic structures on $M$. The action functional of
Green-Schwarz IIB string in certain gauge (so called Schild
gauge) coincides with  a functional defined on $C^{\infty}
(M) \times \Cb^{10|16} \times {\Vc}$ and given by the
formula $S-\mu V$ where $S$ is the functional \action\ and $V$ is
the volume of $M$; both $S$ and $V$ are calculated by means
of symplectic structure $\omega \in {\Vc}$.

Using the remark above one can check easily that the action
functional of Green-Schwarz string can be obtained from the
IKKT Matrix model in the limit $N \rightarrow \infty$.

The proof that the Green-Schwarz string can be described by
means of action functional \action\ requires some work (see \ikkt).
However, almost without calculation one can say that bosonic
part of the action functional $S - \mu V$ leads at the level
of classical equations of motion to the standard bosonic
string. This follows from the remark that the area of embedded
surface (Nambu-Goto action) can be expressed in terms of
Poisson bracket:
$$
\hbox{Area} = \int_M \left( \sum_{i,j} \{ X_i , X_j
\}^2 \right)^{1/2} \omega
$$
where $M$ is a two-dimensional manifold with symplectic
structure corresponding to the 2-form $\omega$. If $\omega =
{\rm const} \cdot d \xi_1 \cdot d\xi_2$ we can identify
equations of motion for Nambu-Goto string with equations of
motion corresponding to a functional we get replacing the
exponent $1/2$ in the expression for the area by any other
number and including an additional term$ -\mu V$.( The new
functional depends not only on fields $ X_i$, but also on
symplectic structure on $M$; symplectic volume of $M$
is denoted by $V$.) In particular,
taking the exponent equal to $1$ we see that bosonic part of
the action $S-\mu V$ leads to the standard equations of motion of
bosonic string.  Therefore the
theory obtained from
this action  can be considered as
supersymmetrization of bosonic string; hence in any case it
leads to a kind of superstring.

One can construct a sequence of maps $\sigma_N : C^{\infty}
(M) \rightarrow {\rm Mat}_N$ where $M$ is a two-dimensional
compact symplectic manifold in such a way that in the limit
$N \rightarrow \infty$
$$
\sigma_N \{ f,g \} - N [\sigma_N (f) , \sigma_N (g) ]
\rightarrow 0
$$
$$
{1 \over N} \int_M f dV - {\rm Tr} \, \sigma_N (f) \rightarrow
0 \, .
$$

The connection between action functionals of IKKT Matrix
model and Green-Schwarz superstring follows immediately from
the existence of maps $\sigma_N$. The maps $\sigma_N$ can be
constructed explicitly when $M$ is a sphere or a torus (see
e.g. [3]).
 The existence of such maps can be derived
from well-known properties of quantization procedure.
Recall, that in semiclassical approximation ($\hbar
\rightarrow 0$) the commutator of quantum observables is
related to the Poisson bracket of classical observables:
\eqn\semicom{
\sigma_{\hbar} \{ f,g \} \approx {1 \over \hbar} \,
[\sigma_{\hbar} (f), \sigma_{\hbar} (g)] \, .
}
Here $\sigma_{\hbar} (f)$ stands for the operator
corresponding to a function $f \in C^{\infty} (M)$ where $M$
is a symplectic manifold. Rigorous construction of the maps
$\sigma_{\hbar}$ can be given in the case when $M$ is a
Kaehler manifold. If $M$ is compact, then the number of quantum
states is finite; in semiclassical approximation it is equal
to
\eqn\semidim{
N = {\hbox{volume} \, (M) \over (2\pi \hbar)^{\dim M/2}} \, .
}
This means that $\sigma_{\hbar} (f)$ can be regarded as $N
\times N$ matrix. Using these remarks and the relation
$$
\int_M f dV \approx (2\pi \hbar)^{\dim M/2} \, {\rm Tr} \,
\sigma_{\hbar} (f)
$$
we obtain  the necessary properties of maps $\sigma_N$
in the case $\dim M =2$.
We also see that the limit $N,R\rightarrow\infty$ should be taken
with $N/R$ fixed to keep the action finite.
The case $\dim M > 2$ also makes sense and
describes higher dimensional objects, branes.

Let us notice that we can weaken the conditions on the Lie
algebra ${\Gc}$, assuming that the inner product is
defined only on its commutant ${\Gc}'$ (i.e. on the
minimal ideal containing all elements of the form $[A,B]$,
$A \in {\Gc}$, $B \in {\Gc}$). The algebra ${\Gc}$
acts on ${\Gc}'$ by means of adjoint representation; we
assume that the inner product on ${\Gc}'$ is invariant
with respect to this action. In this case the expression \action\
still makes sense if $X_0 , \ldots , X_9 \in {\Gc}$,
$\Psi^1 , \ldots , \Psi^{16} \in \Pi {\Gc}'$. All symmetries
of the functional \action\ remain valid in this more general
situation.

The functional \action\ can be obtained from ten-dimensional
SYM theory by means of reduction to a
point (in other words, we restrict the action functional of
this theory to constant fields). It is interesting to notice
that conversely the action functional of SYM theory
on $\Rb^{10}$ is contained in \action\ for the
case when ${\Gc}$ consists of operators acting on
$C^{\infty} (\Rb^{10})$ and having the form $A+B$ where $A$
is a first order differential operator with constant
coefficients and $B$ is an operator of multiplication on a
function decreasing at infinity.

It is easy to verify that the functional \action\ makes sense for
the Lie algebra at hand (one should apply the remark above).

A BPS state is defined as a state that is annihilated
by some of the supersymmetry transformations. Let us consider
 a state
determined by matrices $X_0,...,X_9$ obeying
the condition that all commutators $[X_i,X_j]$ are scalar
matrices.(We assume that $\Psi^{\alpha}=0$.) It is easy to
find linear combinations of supersymmetry transformations
that annihilate such a state and check that one half of
supersymmetries are preserved. Of course, a commutator of
two finite-dimensional matrices cannot be a non-zero
scalar matrix. However, we can consider BPS states
determined by infinite-dimensional matrices; they are
important because we take the limit when the size of
matrices tends to infinity.

\newsec{Toroidal compactification}

Let us first discuss the compactification of IKKT model on a
circle.  We would like to restrict the action functional of this
model to the subspace consisting of points $(X_i ,
\Psi^{\alpha})$ that remain in the same gauge class after a
shift by a real number $2\pi R_0$ in the direction $X_0$. In other
words, we should consider such points that there exists an
invertible matrix $U$ obeying
\eqn\torone{\eqalign{
X_0 + 2\pi R_0 &= U X_0 U^{-1} \ , \ X_i = U X_i U^{-1} \quad
\hbox{for} \quad i > 0 \, , \cr
\Psi^{\alpha} &= U \Psi^{\alpha} U^{-1} .
}}
Taking the trace of
both sides of the first equation we see that finite-dimensional
matrices cannot satisfy these conditions. However, if $X_i$ and
$\Psi^{\alpha}$ are operators in infinite-dimensional Hilbert
space ${\cal H}$ one can easily find solutions of \torone. Let ${\cal
H}$ be a space of functions $f(s)$ depending on the point $s \in
S^1 = \Rb / 2\pi \Zb$ and taking values in Hilbert space ${\cal
E}$ (i.e. ${\cal H} = L^2 (S^1) \otimes {\cal E})$. Then we can
take
\eqn\solone{
\matrix{
X_0 =2\pi iR_0 \, {\partial \over \partial s} + {\cal A}_0 (s) \, , \
X_k = {\cal A}_k (s) \ \hbox{for} \ k > 0 \, , \hfill \cr
\cr
\Psi^{\alpha} = \psi^{\alpha} (s) \, , \ (Uf)(s) = e^{is} f(s)
\, . \hfill \cr
}}
Here ${\cal A}_k (s)$ and $\psi^{\alpha} (s)$ are functions on
$S^1$ taking values in the space of operators acting on ${\cal
E}$; they can be considered in natural way as operators acting on
${\cal H}$.

 We will restrict ourselves to the case when the
operators $X_k$ are Hermitian and the operator $U$ is unitary.
Then the operators
$A_k (s)$ should
be Hermitian for every $s \in S^1$.

One can prove that all other solutions to \torone\ are
unitary equivalent (gauge equivalent) to the solution \solone. To
give a proof we consider the decomposition of ${\cal H}$ into a
direct sum ${\cal H} = \sum_{m \in \Zb} {\cal H}_m$ of
$X_0$-invariant subspaces where the spectrum of the Hermitian operator
$X_0$ restricted
to ${\cal H}_m$ lies in the interval $2\pi m R \leq \lambda < 2\pi
(m+1) R$.

It is clear that $U$ acts from ${\cal H}_m$ into ${\cal
H}_{m+1}$; moreover $U$ can be regarded as an isomorphism of
${\cal H}_m$ and ${\cal H}_{m+1}$. This statement permits us to
identify ${\cal H}$ with a direct sum of infinite number of
copies of ${\cal E} = {\cal H}_0$. In other words, a point of
${\cal H}$ can be considered as an ${\cal E}$-valued function on
$\Zb$ and the operator $U$ corresponds to a shift $m \rightarrow
m+1$. Replacing $\ell_m$, $m \in \Zb$, with a function $\sum_m
\ell_m  e^{ims}$ we obtain a representation of arbitrary
solution to \torone\ in the form \solone.

We should return now to finite-dimensional matrices. As we have
seen only approximate solutions to \torone\ are possible in this case.
To obtain such solutions we assume that ${\cal E}$ has finite
dimension $M$, and replace the differential operator $iR_0 \,
{\partial \over \partial s}$ by a difference operator that tends
to $iR_0 \, {\partial \over \partial s}$ in the limit when lattice
spacing $a$ tends to zero.

Substituting these approximate solutions in the action
functional of the IKKT model and taking $a \rightarrow 0$ we
obtain (after rescaling) an action functional of the form
\eqn\bfssact{\eqalign{
S = & \ {2\over R} \sum_{1 \leq i \leq 9} \int {\rm Tr} (\nabla A_i
(s))^2 ds + \sum_{1 \leq i,j \leq 9} R \int {\rm Tr} [A_i
(s) , A_j (s)]^2 ds \cr
+ & \ 2 \int {\rm Tr} \, \psi^{\alpha} (s) \Gamma_{\alpha
\beta}^0 \nabla \psi^{\beta} (s) ds \cr
+ & \ 2R \sum_{1 \leq i \leq 9} \int {\rm Tr} \,
\psi^{\alpha} (s) \Gamma_{\alpha \beta}^i [A_i (s) ,
\psi^{\beta} (s)] ds
}}
where $\nabla{\varphi} = R(iR_0 \, {\partial \over \partial s}\varphi +
[A_0 , \varphi])$.

This can be regarded as the action functional for matrix quantum
mechanics, with $s$ a compact Euclidean time coordinate.
After Wick rotation, we obtain conventional matrix quantum mechanics,
the starting point for the BFSS model.
One can also say that compactified IKKT model is the BFSS model at
finite temperature, and obtain the non-compactified IKKT model
in the limit when the temperature tends to infinity.

\subsec{Compactification on the standard $T^2$}

Let us consider now a compactification of the IKKT model in two
directions, $X_0$ and $X_1$. This means that we should solve
the equations
\eqn\tortwo{\eqalign{
X_0 + R_0 = & \ U_0 X_0 U_0^{-1} \ , \quad X_1 + R_1 = U_1 X_1
U_1^{-1} \cr
X_i = & \  U_j X_i U_j^{-1} \quad \hbox{if} \quad i \not= j \,
, \ i=0, \ldots ,9 \, , \ j=1,2 \cr
\Psi^{\alpha} = & \ U_j \Psi^{\alpha} U_j^{-1} \, . \cr
}}
Here $R_0$ and $R_1$ are complex constants considered as
scalar operators. We will describe solutions to these
equations where $X_i$, $\Psi^{\alpha}$ and $U_j$ are
operators on an infinite-dimensional Hilbert space ${\cal H}$.

It is easy to derive from \tortwo\ that $U_0 U_1 U_0^{-1}
U_1^{-1}$ commutes with $X_i$ and $\Psi^{\alpha}$. Therefore
it is natural to assume that $U_0 U_1 U_0^{-1} U_1^{-1}$
is a scalar operator, i.e.
\eqn\nctortwo{
U_0 U_1 = \lambda U_1 U_0
}
where $\lambda \equiv e^{2\pi i\theta}$ is a complex constant.

First of all, it is
easy to analyze the case $\lambda = 1$. In this case one can
consider ${\cal H}$ as the space of ${\cal E}$-valued
functions on the torus $S^1 \times S^1$, where ${\cal E}$ is
a Hilbert space, and take $X_0 =i R_0 \, {\partial \over
\partial s_0} + {\cal A}_0 (s_0 , s_1)$, $X_1 =i R_1 \,
{\partial \over \partial s_1} + {\cal A}_1 (s_0 , s_1)$,
$X_i = {\cal A}_i (s_0 ,s_1)$ for $i > 1$, $\Psi^{\alpha} =
\psi^{\alpha} (s_0 ,s_1)$. Here $s_0$, $s_1$ are angle
variables (i .e. $0 \leq s_i < 2\pi$) and ${\cal A}_i$,
$\psi^{\alpha}$ functions on the torus taking values in the
space of linear operators acting on ${\cal E}$. One can
consider instead of ${\cal E}$-valued functions on a torus
sections of a vector bundle $\alpha$ on a torus with typical
fiber ${\cal E}$. Then we should replace $R_0 \, {\partial
\over \partial s_0}$, $R_1 \, {\partial \over \partial s_1}$
with $\nabla_0$, $\nabla_1$ where $\nabla_0$, $\nabla_1$
specify a constant curvature connection and ${\cal A}_i$, $\psi_{\alpha}$
should be considered as sections of a bundle, having as a
fiber over a point $b \in S^1 \times S^1$ the space of
linear operators acting in the corresponding fiber of the
bundle $\alpha$.

One can check that this solution of \tortwo, used
in \refs{\taylor,\bfss,\grt}, is in some sense generic.
The discussion in the previous subsection generalizes to show that
the action functional becomes that for two-dimensional SYM,
and after Wick rotation becomes that for $1+1$ (one space and one time)
dimensional SYM.
Thus the BFSS model compactified on $S^1$ is described by $1+1$
dimensional SYM.

Finally, given a $d$-dimensional solution, we can produce a $d+1$-dimensional
solution which can be used to define the BFSS model compactified on a
$d$-dimensional space, by choosing another matrix coordinate $X^{d+1}$ and
adjoining the relation
$X^{d+1} + R^{d+1} =  \ U_{d+1} X^{d+1} U_{d+1}^{-1}$ where
$U_{d+1}$ commutes with all other $U_i$ and $X^i$.
Thus we can regard any solution to \tortwo\ as also defining a compactification
of the BFSS model on $T^2$.

\subsec{Compactification on noncommutative $T^2$}

We now study the solutions to \tortwo\ and \nctortwo\ for the
case $\lambda \not= 1$.
Let us suppose that $U_0$, $U_1$ are fixed. Then we can
start by finding a particular solution. After that
we will describe the set $E$ consisting of all
operators commuting with $U_0$, $U_1$. The general solution
to \tortwo\ has the form $X_0 = x_0 + A_0$, $X_1 = x_1 + A_1$,
where $(x_0 , x_1)$ is a particular solution, and $A_0 ,A_1 \in
E$. To get the general solution we also take
as $X_i$, $i>1$ arbitrary even elements of $E$, and
as $\Psi^{\alpha}$ arbitrary odd elements of $E$.

Let us consider the space $\Cb^q$ as the space $C(\Zb_q)$ of
functions on finite group $\Zb_q = \Zb / q\Zb$. For every $q
\in \Zb$, $p\in \Zb_q$
(with $p$ and $q$ relatively prime)
we define the operator $W_0$ as the
operator transforming the function $f(k)$ into $f(k-p)$ and
the operator $W_1$ as the operator of multiplication by $\exp
(-2\pi i k/q)$. It is easy to check that
$$
W_0 W_1 = \exp (2\pi i \, p/q) \, W_1 W_0 \, .
$$
We can construct also operators $V_0$ and $V_1$ acting on
the space of smooth functions on $\Rb$ of fast decrease at
infinity as operators transforming a function $f(s)$ into
$e^{ 2\pi i\gamma s} f(s)$ and $f(s+1)$ respectively. These
operators obey
\eqn\ncompart{
V_0 V_1 = e^{- 2\pi i \gamma} V_1 V_0 \, .
}
If $\gamma$ is real, the operators $V_0$ and $V_1$ act on
the Schwartz space $\Hc \equiv S(\Rb)$. They can be considered also
as unitary operators on the space of square-integrable
functions on $\Rb$. In the general case
one can
consider $V_0$ and $V_1$, as operators on the space of
smooth functions that decrease faster than any exponential
function. Now we consider the operators $U_0 = V_0 \otimes
W_0$ and $U_1 = V_1 \otimes W_1$ acting on the space $\Hc_{p,q}$
of functions defined on $\Rb \times \Zb_q$. They obey
\eqn\ucomtwo{
U_0 U_1 = e^{-2\pi i \gamma + 2\pi i p/q} U_1 U_0
}
and thus we have a solution of \nctortwo\ if $\gamma=p/q-\theta$.

One can describe $U_0$ and $U_1$ directly as operators
transforming a function $f(s,k)$ where $s \in \Rb$, $k \in
\Zb_q$, into $e^{2\pi i \gamma s} f(s,k-p)$ and into $\exp
\left( -2\pi i \, {k \over q} \right) f(s+1,k)$
correspondingly. We define the operators $X_0$ and $X_1$ on
this function space, to be denoted ${\Hc_{p,q}}$,
by the formula
\eqn\partsoltwo{\eqalign{
(X_0 f) (s,k) =& \  i \nu \, {\partial f (s,k) \over \partial
s} \cr
(X_1 f) (s,k) =& \ \tau s f(s,k) \, . \cr
}}
It is easy to check that these operators obey \tortwo\ with $R_0 =
2\pi \nu \gamma$ and $R_1 = \tau$.
Their commutator is
\eqn\xcomtwo{
[X^0,X^1] = {i\over 2\pi\gamma} R_0 R_1.
}
This result can also be thought of determining the dimension
of ${\Hc_{p,q}}$, by identifying $\hbar \sim R_0 R_1 / 2\pi\gamma$
and using the semiclassical result \semidim.
This leads to $\dim \Hc = |\gamma|$ and
\eqn\nctwodim{
\dim \, {\Hc}_{p,q} = \dim \Hc \times \dim \Cb^q = |p-q\theta| ,
}
a result we discuss further below.
This will turn out to agree with the notion
of dimension in noncommutative geometry, as we explain
in section 4.

Now we should describe the set $E$ of operators
commuting with $U_0$ and $U_1$. It is easy to verify that
the operator $Z_0$ defined by
$$
(Z_0 f) (s,k) = \exp \left( {2\pi i \over q} s \right) f(s,k-1)
$$
commutes with $U_0$, $U_1$, and the operator $Z_1$
transforming $f(s,k)$ into $e^{i\nu k} f(s+\sigma ,k)$
commutes with $U_0$, $U_1$ if $\sigma ={ 1 \over {\gamma q}} $,
$\nu q= 2\pi a$, $ap+bq=1$, $a$ and $b$ are integers.

Of course, all operators obtained from $Z_0$, $Z_1$ by means
of addition and multiplication and all limits of such
operators commute with $U_0$, $U_1$; one can prove that all
operators commuting with $U_0$, $U_1$ can be obtained this
way. In other words, the algebra $E$ of operators commuting
with $U_0$, $U_1$ is generated by $Z_0$, $Z_1$.
They obey the commutation relation
\eqn\dualtortwo{
Z_0 Z_1 = \lambda' Z_1 Z_0
}
with $\lambda' \equiv e^{2\pi i\theta'}$ and
\eqn\dualtheta{
\theta' = {a \theta +b \over p - q \theta }
}
where $a$ and $b$ are integers obeying $ap +bq =1$.
Thus we can think of $A_0$, $A_1$ and the remaining
$X^i$ and $\Psi$ as fields on the dual noncommutative torus
with parameter $\theta'$.

\subsec{Dimension formula via explicit limiting construction}

We can think of the solutions of the previous section as obtained
by taking the limit of a finite dimensional matrix construction.  Besides
being slightly more concrete, this will allow us to justify the
formula \nctwodim.  For more detail on such constructions see \schwarz.

To find finite dimensional approximate solutions of \tortwo,
we can replace the function space on which the operators $V_i$ act
with a lattice approximation $C(\Zb_M)$.
So, we introduce a lattice spacing $a$ and
replace the variable $s$ with $an$ where $n=0,\dots,M-1$.
The operators $V_i$ can then be represented by ``clock and shift''
operators isomorphic to $W_i$ but acting on $C(\Zb_M)$.
In the limit $M\rightarrow\infty$, these will be able to approximate
\ncompart\ with an arbitrary $\gamma$.

The limit $M\rightarrow\infty$ is then taken with a simultaneous
rescaling of the trace to keep it finite.
The point is that this rescaling should be determined by ``local''
considerations, meaning local either in ``index space'' or in the
resulting string or membrane world-volume theory.
Such a local rescaling should depend on the lattice spacing and the
volume of the world-sheet; for example we can have
$$\Tr = {a\over {\rm volume}} \Tr_{\rm original} .$$
Writing
$$\eqalign{
(X_0 f)(n,k) &= {i\over a}\left(f(n+1,k)-f(n,k)\right) \cr
(X_1 f)(n,k) &= a n\ f(n,k) ,
}$$
and comparing with \partsoltwo\ determines the
lattice spacing $\nu \sim 1/a$ and ${\rm volume} = M a = \tau$.
This determines a rescaled trace
$$\Tr = {1\over \nu\tau} \Tr_{\rm original}
= {2\pi\gamma\over R_0R_1} \Tr_{\rm original} .$$
Now the factors ${2\pi/ R_0R_1}$ are local and so it is a matter
of convention whether we keep them in the definition,
but the factor $\gamma$ is not.  We will drop the $2\pi$ and associate
the factor $1/R_0R_1$ with the volume of the two-torus,
$1/R_0R_1 = \int_{T^2} 1$.
This determines the final trace
$$\Tr_{\rm final} = \gamma\ \Tr_{\rm original} $$
and the dimension $\dim \Hc \sim \Tr 1 \sim \gamma$.
Finally,
taking the tensor product with the space $C(\Zb_q)$ on which the $U_i$
act leads to the formula \nctwodim.

It is important to notice that we can construct
many other approximate solutions adding terms that can be neglected
when we remove the cutoff; they don't change the action functional
we obtained. However, due to ultraviolet divergences they can contribute
to physical quantities. If the action functional of the compactified theory is
non-renormalizable, one expects that the contribution of other
approximate solutions can be described in terms of additional fields
arising in the theory. This remark may give an explanation of the origin
of the new fields found in \tfour.

\subsec{Compactification on $T^d$}

This discussion generalizes directly to the multidimensional
torus as follows.
Let $\tau^{(1)} , \ldots , \tau^{(d)}$ denote $d$ linearly
independent ten-dimensional vectors. We should find operators
$U_1 , \ldots , U_d$ obeying
\eqn\nctord{
U_i U_j = \lambda_{ij} U_j U_i
}
and operators $X_0 , \ldots , X_9$ obeying
\eqn\tord{
X_i + \tau_i^{(k)} = U_k X_i U_k^{-1} \, .
}
We restrict ourselves to the most interesting case when
$\tau^{(k)}$ are real, $X_i$ are Hermitian and $U_i$ are unitary.

One can find solutions to these equations in the following way.
Let us consider an abelian group $\Gamma$ that can be
represented as a direct sum of groups $\Rb$, $\Zb$ and $\Zb_m$.
Let us fix $d$ elements $a_1 , \ldots , a_d$ of the group
$\Gamma$ and $d$ characters $\chi_1 , \ldots , \chi_d$ of the
group $\Gamma$. (We consider $\Gamma$ as a group with respect to
addition, therefore a character can be defined as complex valued
function on $\Gamma$ obeying $\chi (\gamma_1 + \gamma_2) = \chi
(\gamma_1) \chi (\gamma_2)$, $\vert \chi (\gamma) \vert =1$.) We
can define operators $U_1 , \ldots , U_d$ by the formula
$$
(U_i f) (\gamma) = \chi_i (\gamma) f(\gamma + a_i) \, .
$$
These operators act on the space $S(\Gamma)$ consisting of
functions on $\Gamma$ that tend to zero at infinity faster than
any power. It is easy to check that $U_1 , \ldots , U_d$ obey \nctord\ with
$$
\lambda_{ij} = {\chi_i (a_j) \over \chi_j (a_i)} \, .
$$
Now we can define operators $X_k$ in the following way:
$$
(X_k f) (s,g) = A_k^i s_i f(s,g) + B_{ki} \,
{\partial f (s,g) \over \partial s_i} \, .
$$
Here we represent $\Gamma$ as $\Rb^m \times \Gamma'$, where
$\Gamma'$ is a discrete group, $s=(s_1 , \ldots , s_m)$ $\in
\Rb^m$, $g \in \Gamma'$. It is easy to check that $X_0 ,
\ldots , X_9$ obey \tord\ with
$$
\tau_j^{(k)} = A_j^i a_{ki} + \alpha_k^i B_{ij}
$$
where $a_{ki}$ stands for the $i$-th component of the projection
of $a_k \in \Rb^m \times \Gamma'$ onto $\Rb^m$ and $\alpha^{ik}$
is defined by the formula
$$
(\chi^k)^{-1} \, {\partial \chi^k \over \partial s_i} = \alpha_k^i
\, .
$$
The commutator $[X_k , X_j]$ is a scalar operator (a
$c$-number), therefore $X_0 , \ldots ,$ $X_9$ determine a BPS
state.

To find other solutions of \tord\ we should describe the algebra $E$
of operators commuting with $U_1 , \ldots ,U_d$. It is easy to
check that an operator $Z$ transforming $f(\gamma)$ into $\beta
(\gamma) f (\gamma + b)$ commutes with operators $U_1 , \ldots ,
U_d$ of $\chi_i (b) = \beta (a_i)$, $i=1,\ldots ,d$. Under
certain conditions one can prove that the algebra $E$ is
generated by operators $Z$ of this form.

\newsec{Noncommutative geometry approach}

Noncommutative geometry starts from the duality of a space with its
algebra of functions:
knowing
the structure of the associative commutative algebra $C(X)$
of complex-valued continuous functions on topological space
$X$ we can restore the space $X$. This means that all
topological notions can be expressed in terms of algebraic
properties of $C(X)$. For example, vector bundles over
compact space $X$ can be identified with projective modules
over $C(X)$. (By definition a projective module is a module
that can be embedded into a free module as a direct summand.
Talking about modules we have in mind left modules. We
consider only finitely generated modules. The space of
continuous sections of vector bundle over $X$ can be
regarded as a $C(X)$-module; this module is projective.)

It was shown that one can introduce many important geometric
notions and prove highly non-trivial theorems considering an
associative algebra ${\cal A}$ as the noncommutative analog of
topological space. For example, a vector bundle is by
definition a projective module over ${\cal A}$ and one can
develop a theory of such bundles generalizing the standard
topological theory. In particular, one can introduce a
notion of a connection, containing as a special case the
standard notion. We will not give the most general
definition of connection but restrict ourselves to the case
when the algebra ${\cal A}$ is considered together with a
Lie algebra ${\Gc}$ of derivations of ${\cal A}$; the
generators of ${\Gc}$ will be denoted by $\alpha _1 , \ldots
,\alpha _d$. If ${\Hc}$ is a projective module over
${\cal A}$ (``a vector bundle over ${\cal A}$'') we define a
connection in ${\Hc}$ as a set of linear operators
$\nabla_1 , \ldots , \nabla_d$ acting on ${\Hc}$ and
satisfying
$$
\nabla_i (a \varphi) = a \nabla_i (\varphi) + \alpha _i (a) \varphi
$$
(here $a \in {\cal A}$, $\varphi \in {\Hc}$, $i=1,\ldots
,d$). In the case when ${\cal A}$ is an algebra of smooth
functions on $\Rb^d$ or on the torus $T^d$ we obtain the
standard notion of connection in a vector bundle. (The
abelian Lie algebra ${\Gc} = \Rb^d$ acts on $\Rb^d$ or
$T^d$ and correspondingly on ${\cal A}$ by means of
translations.) If $\nabla_i$ and $\nabla'_i$ are two
connections then the difference $\nabla'_i - \nabla_i$
commutes with multiplication by $a$; i.e. $\nabla'_i -
\nabla_i$ belongs to the algebra $E= \hbox{End}_{\cal A}
{\Hc}$ of endomorphisms of the ${\cal A}$-module ${\cal
H}$. It is easy to check that
$$
F_{ij} = \nabla_i \nabla_j - \nabla_j \nabla_i - f_{ij}^k
\nabla_k
$$
where $f_{ij}^k$ are structure constants of ${\Gc}$ also
belongs to $E$. It is clear that $F_{ij}$ should be
considered as a curvature of the connection $\nabla_i$.

Let us specify the notions above for the case when ${\cal
A}$ is $d$-dimensional noncommutative torus, i.e. an
algebra $T_C$ with generators $U_a$ satisfying relations
$$
U_a U_b = C_{ab} U_b U_a \, .
$$
(Here $a,b = 1,\ldots ,d$, $C_{ab}$ are complex numbers,
$C_{ab} = C_{ba}^{-1}$.) In the case when $\vert C_{ab}
\vert =1$ the algebra $T_C$ can be equipped with an
antilinear involution $*$ obeying $U_a^* = U_a^{-1}$ (i.e.
${\cal A}$ is a $*$-algebra). The name ``noncommutative
torus'' is used also for various completions of $T_C$; at
this moment we don't fix a specific completion. (One can say
that different completions specify  different classes of
``functions'' on the same noncommutative space.)
Let us fix an abelian Lie algebra ${\Gc}$ of
automorphisms of $T_C$ generated by operators $\alpha_1 ,
\ldots , \alpha_d$ given by the formula $\alpha_k (U_a) =
2\pi i U_a$ if $k=a$, $\alpha_k (U_a) = 0$ if $k \not= a$.
Then a connection in a module ${\Hc}$ over $T_C$ is
determined by a set of operators $\nabla_1 , \ldots ,
\nabla_d$ in ${\Hc}$ obeying
$$
\nabla_i U_j - U_j \nabla_i = \delta_{ij} U_i \cdot 2\pi i
\, .
$$
In other words, taking $-i \nabla_k = X_k$,
we find a solution to the equation \tord,
defining a toroidal compactification of Matrix theory.

We see that the classification of toroidal compactifications can
be reduced to a problem studied in noncommutative
geometry, and treated in detail in \connesmath, where
proofs of the following statements can be found.
First of all one should fix a
module ${\Hc}$ over the noncommutative torus; we restrict
ourselves to projective modules. Then we should find the
endomorphism algebra $E = \hbox{End}_{T_C} {\Hc}$ of the
module ${\Hc}$, and construct one connection $\nabla_1 ,
\ldots , \nabla_d$. After that the general solution to the
equations
$$
X_i+2\pi i\delta_{ij}=U_j X_i U_j^{-1}, \quad 1\leq i,j \leq d,
$$
$$
X_k = U_j X_k U_j^{-1}, \quad d<k \leq 10,
$$
$$
\Psi^{\alpha}=\psi^{\alpha}
$$
can be written in the form
$$
\eqalign{
&X_a = i \nabla_a + {\cal A}_a \, , \quad 1 \leq a \leq d \,
, \cr
&X_k = {\cal A}_k \quad d < k \leq 10 \cr
&\Psi^{\alpha} = \psi^{\alpha} \, . \cr
}
$$
Here ${\cal A}_s$, $1 \leq s \leq 10$ are arbitrary elements
of $E$, $\psi^{\alpha}$ are arbitrary elements of $\Pi E$.

A fairly complete mathematical theory of projective modules
and of connections on these modules exists for the case when
noncommutative torus $T_C$ is a $*$-algebra (i.e. in the
case $\vert C_{ab} \vert = 1$. In this case it is natural to
consider a completion $T_C^{\infty}$ of the algebra $T_C$
generated by $U_1 , \ldots ,U_d$ including power series
$$
U = \sum C_{\alpha_1 \ldots \alpha_d} U_1^{\alpha_1} \ldots U_d^{\alpha_d}
$$
where the coefficients $C_{\alpha_1 \ldots \alpha_d}$ tend to zero
faster than any power of $\vert \alpha \vert = |\alpha_1| +
\ldots + |\alpha_d|$ as $\vert \alpha \vert \rightarrow
\infty$. One can construct a trace on $T_C^{\infty}$ by the
formula ${\rm Tr} \, U = C_{0,\ldots ,0}$; this trace is
invariant with respect to the Lie algebra ${\Gc}$ of
automorphisms of $T_C^{\infty}$. If we consider projective
modules over $*$-algebra ${\cal A}$ it is natural to equip
such modules with Hermitian metric, i.e. with ${\cal
A}$-valued positive-definite Hermitian inner product
$\langle \ , \ \rangle_{\cal A}$ obeying
$$
\langle \xi ,\eta \rangle_A^* = \langle \eta , \xi \rangle_A
\, , \ \langle \xi ,a\eta \rangle_A = a \langle \xi ,\eta
\rangle_A \, .
$$
It can be proven that such an inner product always exists.
One can introduce a notion of a connection $\nabla_i$
compatible with Hermitian metric requiring that
\eqn\compcon{
\langle \nabla_i \xi ,\eta \rangle_{\cal A} + \langle \xi ,
\nabla_i \eta \rangle_{\cal A} = \alpha_i (\langle \xi ,\eta
\rangle_{\cal A}) \, .
}
The algebra $T_C^{\infty}$ is equipped with a trace ${\rm Tr} \,
\xi$ obeying ${\rm Tr} \, \xi^* = \overline{{\rm Tr} \, \xi}$,
hence we can introduce a complex valued Hermitian inner
product on the module taking
$$
\langle \xi , \eta \rangle = {\rm Tr} \langle \xi ,
\eta \rangle_A \, .
$$
Then it follows from \compcon\ that $\nabla_i$ is a skew-adjoint
operator. (We use the invariance of the trace with respect
to ${\Gc}$.)

If $U_1 , \ldots ,U_d$ are unitary operators in Hilbert
space ${\Hc}$ and $U_i U_j = \lambda_{ij} U_j$ $U_i$ where
$\lambda_{ij}$ are complex numbers, then $\vert \lambda_{ij} \vert =1$.
It is easy to check that $U_1 , \ldots ,U_d$ determine a
module over $T_C^{\infty}$; it follows from the statements
above that in the case where this module is projective,
there exist self-adjoint operators $X_1 , \ldots, X_d$
obeying \tord. It remains to find the algebra $E$ of
endomorphisms of the module ${\Hc}$ to get a description
of general solution of \nctord, \tord. Under certain conditions
one can prove that $E$ is isomorphic to a noncommutative
torus $T_{C'}$ that is dual to the original torus $T_C$
in some sense. In more complicated situations one can get as
$E$ an algebra of matrices with entries from the noncommutative
torus. However, in any case the algebra $E$ has a trace.
One can prove that many of properties of the algebra $E$
are the same as for torus $T_C$ . (More precisely, these two
algebras are strongly Morita equivalent.) If
operators $\nabla_1 , \ldots ,\nabla_d$ constitute a
connection on module ${\Hc}$ then it is easy to check
that for every $e \in E$ we have $[\nabla_a , e] \in E$ and
${\rm Tr} [\nabla_a , e] = 0$ (i.e. the trace on $E$ is
invariant with respect to the natural action of the
connection on $E$). Using this remark and the fact that
$[\nabla_a , \nabla_b ] = F_{ab} \in E$ we see that the
functional \action\ can be considered as a functional on
the space of all connections; we interpret this functional as
the action functional of the compactified theory.

The curvature of a connection in our case when ${\Gc}$ is
an abelian Lie algebra takes the form $F_{ij} = [\nabla_i ,
\nabla_j]$. These means that connections with constant
curvature (i.e. connections with $F_{ij} = \gamma_{ij} \cdot
1$, where $\gamma_{ij}$ are constants) correspond to
BPS states. Connections with constant curvature exist for
every projective module if $d=2$.

In the case of $d=2$ an explicit description of all
projective modules over $T_C^{\infty}$ and connections in
these modules was given in \connesmath. The preceding section contains
basically a translation of some results of this paper into a
simpler language. Namely, the formula \ucomtwo\ shows that the
operators $U_0$, $U_1$ determine a module over a
noncommutative torus $T_C$ with $C_{01} = \exp (-2\pi i
\theta)$ where $\theta = \gamma   - p/q$,
$\theta$ and $\gamma$ are real numbers, $p$ and $q$ are
relatively prime integers. We will denote this module by
${\Hc}_{p,q}^{\theta}$; one can prove that it is
projective. We can consider also a projective module $({\cal
H}_{p,q}^{\theta})^n$ consisting of $n$ copies of
${\Hc}_{p,q}$. Algebra $E_n$ of endomorphisms of the
module $({\Hc}_{p,q}^{\theta})^n$ is isomorphic to
the algebra ${\rm Mat}_N (E)$ of matrices with entries from
the algebra $E$ of endomorphisms of ${\cal
H}_{p,q}^{\theta}$. We stated already that algebra $E$
is generated by operators $Z_0$, $Z_1$; therefore it is
isomorphic to noncommutative torus $T_{C'}$ where
$$
C' = \exp \left( 2\pi i \, {a \theta +b \over p - q \theta }
\right) \, ,
$$
$a$ and $b$ are integers obeying $ap +bq =1$. The operators
$\nabla_0 = -i X_0 + \alpha_0$, $\nabla_1 = -i X_1 +
\alpha_1$ where $X_0$, $X_1$ are defined in \partsoltwo, $\alpha_0$,
$\alpha_1$ are real numbers, determine a compatible
connection with constant curvature in ${\cal
H}_{p,q}^{\theta}$. Obvious (``block diagonal'')
construction gives similar connections in $({\cal
H}_{p,q}^{\theta})^n$; it is proved in \connesmath\ that every
compatible connection with constant curvature in $({\cal
H}_{p,q}^{\theta})^n$ can be obtained this way (up to
gauge equivalence) and that moduli space of such connections
with respect to gauge equivalence can be identified with
$(T^2)^n / S_n$. Here $T^2$ is two-dimensional torus and the
symmetric group $S_n$ acts in standard way on $(T^2)^n$.

In the case when $\lambda = \exp (-2\pi i \theta)$ and
$\theta$ is irrational, free modules and modules $({\cal
H}_{p,q}^{\theta})^n$ exhaust all projective modules
over noncommutative torus $T_C^{\infty}$. In the case when
$\theta$ is rational there are additional projective
modules, also described in \connesmath. We will not repeat this
description; however, it is interesting to mention that the
identification of the moduli space of compatible connections
having constant curvature with $(T^2)^n / S_n$ remains valid in
all cases.

The usual notion of dimension of a vector bundle extends to
projective modules over the NC torus, but it is no longer
necessarily an integer or even a rational number. With the above
notations one has
$$
\dim \, {\Hc}_{p,q} = \vert p - q \, \theta \vert \, .
$$
This is obtained in two equivalent ways. The first one is to write
the  projective module as the range of a projection $P$ belonging to
the $q\times q$ matrices over the algebra, the trace of $P$ is then
well defined
and independent of any choice, provided the trace on the algebra is equal
to $1$ on the unit element.  The second one is to count the least number
$l(N)$ of generators of the direct sum of $N$ copies of the module over the
weak closure of the algebra.  The dimension is then the limit of the
ratio $l(N)/N$.

The $K$ theory group which classifies  projective modules is
the rank two abelian group $\Zb^2$. Since its elements are classes
of virtual projective modules (i.e. formal differences of classes of
f.p.-modules) it has a natural ordering, whose cone of positive
elements is the set of classes of actual f.p.-modules. The
corresponding cone in $\Zb^2$ is
$$
\{ (x,y) \in \Zb^2 \, ; \, x - \theta y > 0 \}
$$
and the coordinates of the module ${\Hc}_{p,q}$ are
$$
x = \pm p \ , \ y = \pm q
$$
where $\pm = {\rm sign} \, (p-q \, \theta)$.

Even though these modules do not in general have integral dimension,
the integral curvature $\Tr F$ is independent of the choice of connection
and remains quantized. The reason behind this fact is that the integral
curvature computes the index of a Fredholm operator  (cf \connes).

\subsec{Moduli space and duality for $\Tb^\theta$}

We shall now describe the natural moduli space
(or more precisely, its covering Teichmuller space) for the
noncommutative tori, together with a natural action of $SL(2,\Zb)$
on this space. The discussion parallels the description of
the moduli space of elliptic curves but we shall find that our moduli space
is the boundary of the latter space.

We first observe that as the parameter $\theta \in \Rb
/ \Zb$ varies from 1 to 0 in the above construction of ${\cal
H}_{p,q}^{\theta}$ one gets a monodromy, using the isomorphism $\Tb_{\theta}^2
\sim \Tb_{\theta + 1}^2$. The computation shows
that this monodromy is given by the transformation $\left[ \matrix{1 &-1 \cr 0
&1 \cr} \right]$ i.e., $x \rightarrow x-y$, $y \rightarrow y$ in terms of the
$(x,y)$ coordinates in the $K$ group. This shows that in order to
follow the $\theta$-dependence of the $K$ group, we should consider the algebra
${\cal A}$ together with a choice of isomorphism,
$$
K_0 ({\cal A}) \build{\simeq}_{}^{\rho} \Zb^2 \ , \ \rho \,
(\hbox{trivial module}) = (1,0) \, .
$$
Exactly as the Jacobian of an elliptic curve appears as a quotient
of the $(1,0)$ part of the cohomology by the lattice of integral
classes, we can associate canonically to ${\cal A}$ the following
data:

\item{1)}
The ordinary two dimensional torus $\Tb = HC_{\rm even} ({\cal A}) / K_0 ({\cal
A})$  quotient of the cyclic homology of ${\cal A}$ by the image of
$K$ theory under the Chern character map,

\item{2)}
The foliation $F$ (of the above torus) given by the natural filtration of
cyclic
homology (dual to the filtration of $HC^{\rm even}$ ).

\item{3)}
The transversal $T$ to the foliation given by the geodesic
joining $0$ to the class $[1] \in K_0$ of the trivial bundle.

It turns out that the algebra associated to the
foliation $F$, and the transversal $T$ is isomorphic to ${\cal A}$, and that a
purely geometric construction associates to every element $\alpha
\in K_0$ its canonical representative from the transversal given by the
geodesic joining $0$ to $\alpha$.
(Elements of the algebra associated to the transversal $T$ are
just matrices  $a(i,j)$ where the indices $(i,j)$ are arbitrary pairs of
elements $i,j$ of $T$ which belong to the same leaf. The algebraic rules are
the same as for ordinary matrices.  Elements of the module associated to
another transversal $T'$ are rectangular matrices, and the dimension of the
module is the transverse measure of $T'$)

This gives the above description of the modules ${\Hc}_{p,q}$
(where in fact the correct formulation uses the Fourier transform of $f$ rather
than $f$).
The above is in perfect analogy with the isomorphism of
an elliptic curve with its Jacobian. The striking difference is
that we use the {\it even} cohomology and $K$ group instead of the
odd ones.

It shows that, using the isomorphism $\rho$, the whole
situation is described by a foliation $ dx= \theta dy$ of $\Rb^2$ where
the exact value of $\theta$ (not only modulo 1) does matter now.

Now the space of translation invariant foliations of
$\Rb^2$ is the boundary $N$ of the space $M$ of translation
invariant conformal structures on $\Rb^2$, and with $\Zb^2 \subset
\Rb^2$ a fixed lattice, they both inherit an action of $SL(2,\Zb)$.
We now describe this action precisely in terms of the pair $({\cal
A} , \rho)$. Let $g = \left[ \matrix{ a &b \cr c &d \cr} \right]
\in SL(2,\Zb)$. Let ${\cal E} = {\Hc}_{p,q}$ where $(p,q) = \pm
(d,-c)$, we define a new algebra ${\cal A}'$ as the commutant of
${\cal A}$ in ${\cal E}$, i.e. as
$$
{\cal A}' = {\rm End}_{\cal A} ({\cal E}) \, .
$$
It turns out (this is called Morita equivalence) that there is a
canonical map $\mu$ from $K_0 ({\cal A}')$ to $K_0 ({\cal A})$
(obtained as a tensor product over ${\cal A}'$) and the isomorphism
$\rho' : K_0 ({\cal A}') \simeq \Zb^2$ is obtained by
$$
\rho' = g \circ \rho \circ \mu \, .
$$
This gives an action of $SL (2 , \Zb)$ on pairs $({\cal A} , \rho)$
with irrational $\theta$ (the new value of  $\theta$  is $(a \theta +b)/(c
\theta +d)$
and for rational values one has to add a
point at $\infty$).

Finally another group $SL(2,\Zb)$ appears when we discuss
the moduli space of flat metrics on $\Tb_{\theta}^2$. Provided we
imitate the usual construction of Teichm\"uller space by fixing an
isomorphism,
$$
\rho_1 : K_1 ({\cal A}) \rightarrow \Zb^2
$$
of the {\it odd} $K$ group with $\Zb^2$, the usual discussion goes
through and the results of \connesmath\ show that for all values of $\theta$
one has a canonical isomorphism of the moduli space with the
upper half plane $M$ divided by the usual action of $SL (2,\Zb)$.
Moreover, one shows that the two actions of $SL(2,\Zb)$
actually commute. The striking fact is that the relation between
the two Teichmuller spaces,
$$
N = \partial \, M
$$
is preserved by the diagonal action of $SL(2,\Zb)$.

\newsec{Gauge theory on the noncommutative torus}

An interesting open problem is to classify all gauge theory Lagrangians
admiting maximal supersymmetry, i.e. $16$ real supersymmetries.
It has been shown that
the only such Lagrangians with terms having at most two derivatives are
the dimensional reductions of
ten-dimensional super Yang-Mills \baake.

A different example with higher derivative terms is the Born-Infeld
generalization of SYM which appears as the world-volume Lagrangian of
$N$ parallel D-branes.  Although the Lagrangian is known explicitly
only for gauge group $U(1)$, the type \I\ superstring
theory provides an implicit definition for all classical groups
(see \nbi\ for a recent discussion).

Gauge theory on the noncommutative torus
provides another example
(although, one which loses Lorentz invariance).
As in section 3, we can regard certain solutions of \nctortwo\
or \nctord\ as
defining continuous field configurations on the noncommutative torus.
Specializing the SYM action \action\ to these configurations defines a
generalization of the $d+1$-dimensional SYM action.
This type of construction was first made in \connesmath\ and
is fairly well known in the physics literature
following the work \dhn. 
The supermembrane theory of \dhn\ is the $2+1$ dimensional case,
in the limit in which the Moyal bracket becomes the Poisson bracket.
The bosonic truncation of the Moyal bracket Lagrangian 
has been considered in \hoppe.  
(See also \refs{\zfc,\ncfields}\ and references there for recent related work).

Let us make the construction more concrete in the particular case
of the manifold $\Tb_2^\theta$.
We will do this for an irreducible module, leading to a ``$U(1)$''
connection described by a single one-form on $T^2$,
but all of the definitions can be applied with matrix-valued fields
as well.
As discussed in section 3, the general solution of \nctortwo\
for the fields $A,X$ and $\Psi$ can be expressed as a sum of the
particular solution \partsoltwo\ with a general element of the
algebra $\Tb_2^{\theta'}$.
If we choose an identification of elements of this algebra with
functions on $T^2$, we can write a conventional gauge theory Lagrangian
in terms of these functions.

Let us set $R_1=R_2=2\pi$ for simplicity and choose
the identification
\eqn\identtwo{\eqalign{
A \in T_2^{\theta'} &\rightarrow f_A(\sigma_1,\sigma_2) \cr
Z_1^m Z_2^n &\rightarrow e^{i(m\sigma_1 + n\sigma_2-\pi\theta' mn)}.
}}
The phase factor is present to simplify the reality condition,
$$
A = A^\dagger \rightarrow f_A = f_A^\dagger = f_A^*
$$
(with the last equality for $U(1)$).
We take the index $i=1,2$, while $0 \le \mu \le 9$ is the original $SO(10)$
vector index.
The constant curvature connection
$$
[X^i,X^j] = f_{ij} \cdot {\bf 1}
$$
acts as the elementary derivation,
\eqn\conntwo{
\nabla_i A = [X^i,A] \rightarrow {\p f_A(\sigma_1,\sigma_2)\over\p\sigma_i} .
}
The trace on $\Tb_2^{\theta'}$ is simply represented by
\eqn\tracetwo{
\Tr f = \int d\sigma_1 d\sigma_2\ f(\sigma_1,\sigma_2)
}
while multiplication is represented by the star product
\eqn\multtwo{\eqalign{
AB \rightarrow
(f_A * g_B)(\sigma)
&= \exp\left(
	\pi i\theta'\epsilon^{ij}{\p\over\p\sigma'_i}{\p\over\p\sigma''_j}
  \right)
 f(\sigma') g(\sigma'')
\bigg|_{\sigma'=\sigma''=\sigma} .
}}
The action \action\ depends only on the commutator
\eqn\commtwo{
[A,B] \rightarrow f_A * g_B - g_B * f_A \equiv \{f,g\}_{\theta'}
}
where $\{f,g\}_{\theta'}$ is related to the usual Moyal bracket
as $\{f,g\}_{\theta'} = i\theta'\{f,g\}_{M.B. \hbar=\theta'}$.

Thus we can write the curvature as
\eqn\uonecurv{\eqalign{
F_{\mu\nu} &= f_{\mu\nu} + 
\p_\mu A_\nu - \p_\nu A_\mu + \{A_\mu,A_\nu\}_{\theta'} \cr
&= f_{\mu\nu} + \p_\mu A_\nu - \p_\nu A_\mu
+ {2\pi i\theta'} \left(\p_1 A_\mu \p_2 A_\nu
				- \p_1 A_\nu \p_2 A_\mu \right) + \ldots
}}
and covariant derivative as
\eqn\uonecov{
D_\mu\phi \equiv \p_\mu \phi + \{A_\mu,\phi\}_{\theta'} .
}

Introducing a gauge coupling constant
and adding the term \topterm, the bosonic action is
\eqn\uoneact{\eqalign{
S = {1\over g_{YM}^2}\int d^2\sigma\ &\sum_{\mu,\nu}
F_{\mu\nu}^2 + \gamma_{\mu\nu} F_{\mu\nu} \cr
}}
It enjoys the gauge invariance
\eqn\uonegauge{
\delta A_\mu = \p_\mu \epsilon + i\{\epsilon,A_\mu\}_{\theta'} .
}
Adding fermions, minimally coupled using the covariant
derivative \uonecov, produces a maximally supersymmetric action with
the supersymmetry \susy.

We can generalize the construction to $p+1$ dimensions and
as long as $\theta_{0i}=0$, these are Lagrangians with two time
derivatives which admit a conventional canonical formulation and
canonical quantization.  Even $\theta_{0i}\ne 0$ looks formally
sensible in the context of functional integral quantization.

It is an important question
whether the higher dimensional theories are renormalizable;
whether we have listed all the renormalized couplings;
and whether they actually respect the maximal supersymmetry.
Of course from a mathematical point of view this is still a conjecture
even for conventional SYM, and we will address this question elsewhere.
Let us make two comments, however, supporting the idea that these
theories could be renormalizable in dimensions $p\le 3$ (just as in the
conventional case).

First, for rational $\theta$, these theories are equivalent to
particular sectors in the standard renormalizable $U(N)$
gauge theories.
To the extent that observables are continuous in $\theta$
(which should not be taken for granted), this is already a strong argument.

Second, for general $\theta$, perturbation theory based on the action
\uoneact\ and its matrix generalization is very similar to conventional
gauge perturbation theory, with the main difference being additional factors
such as $\exp i\theta^{\prime\,ij} k_i k_j'$ in the interaction vertices.
The presence of the $i$ in the exponent leads to significant differences
with general higher derivative field theory and
indeed the oscillatory nature of these factors
make the sums over loop momenta {\it more}
convergent than in conventional gauge theory.

We also note that these theories are non-local\footnote*{
in the usual sense; it may well emerge that they are local in some
modified sense.}
without any preferred
scale (the parameter $\theta$ is dimensionless).  This shows up
in the leading (tree level) scattering of a particle from a plane wave
background.  Thus they would not arise as low energy limits
of local field theory, and this is why they have not played a
major role in physics so far.
However, there is no known reason why this should disqualify them
from use in matrix theory.

We finally note that the action \uoneact\
can be generalized to a general (curved) manifold with a metric
and a Poisson structure, by replacing $f * g$ with
the star product of deformation quantization \defquant.
Generally speaking, the result
should be an action derived along the lines of \connesym.
Thus the new parameters in this type of compactification are quite generally
the additional choice of a Poisson structure,
or equivalently (in the non-degenerate case) a closed two-form $\theta$.

\subsec{Conserved charges and energies of BPS states}

In section 6 we will use this construction on $\Tb_\theta^2\times \Rb$
to define the BFSS model on the non-commutative torus.
We will use the quantization of the conserved charges and the
energies of the corresponding BPS states.

This action can be obtained following the discussion in section 2 with
compact $X^0$ and $\theta_{0i}=0$, and Wick rotating $X^0$ to a time
coordinate $t$ on $\Rb$;
The only changes
to \uoneact\ are to make the fields depend on $t$ as well and to include
$\p/\p t$ terms in \uonecurv, \uonecov and \uonegauge.

The conserved quantities in $p+1$ gauge theory on $T^p$ all have
analogs here.  There are the total electric flux
\eqn\totalelf{
e_i = \int d^px\ \Tr \p_0 A_i ,
}
and the total magnetic flux
\eqn\totalmnf{
m_{ij} = {1\over 2\pi}\int d^px\ \Tr F_{ij} .
}

There is a conserved stress tensor which can be derived by the usual
Noether procedure, or by evaluating the conventional gauge theory
stress tensor
$$T_{\mu\nu} = g^{\lambda\sigma} \Tr F_{\mu\lambda}F_{\nu\sigma}
		- {1\over p+1} g^{\mu\nu}\Tr F^2
$$
on the configurations.
This leads to the conserved momenta
$$
P_i = \int d^px\ \Tr \sum_j \p_0 A_j 
	\left(F_{ij} + \gamma_{ij}\right).
$$
We could rewrite it using \totalelf\ and \totalmnf\ as
$$
P_i = \sum_j (m_{ij}+\gamma_{ij}) e^j + P'_i
$$
where $P'_i$ is the contribution from non-constant modes of the fields.

If one considers a state of definite charge,
and adiabatically varies the parameters of the theory, it is possible that
the conserved quantities which remain fixed are not the naive charges
but instead linear combinations depending on the parameters.
For example, the action for SYM with $p=3$ has an additional topological term
$b \int F \wedge F$, and it is known that the charge which is fixed
under variations of $b$ is
$E_i \equiv e_i + b \epsilon_{ijk} m^{jk}$ \witten.
It is this charge which enters into the energy formula for a BPS state.

We will need the analogous statements for this theory.
Without a precise definition of the quantum theory,
they will be somewhat conjectural.
The assumption we will make is that $m_{ij}$ and the total momentum
$P_i$ remain fixed.
Since these are quantized even in the classical theory,
it seems very plausible that they remain fixed under deformation.

The arguments of section 3 leading to the relations \xcomtwo\
and \nctwodim,
$$\eqalign{
\Tr 1 &= \dim {\Hc}_{p,q}^{\theta} = |p-q\theta|; \cr
f_{12} &= {2\pi q V\over (p-q\theta)}
}$$
imply that these are the correct normalization and
flux quantization conditions in this gauge theory.
Here $V=\det g$ where
$g_{ij}$ is the metric on the moduli space of flat connections,
generalizing slightly
the discussion of section 3 where $g_{ii}=R_i^2$ and $g_{ij}=0$.

For integral $\theta$, they reduce to the standard
conditions on the commuting torus.
For example, $(p,q,\theta)=(0,1,N)$
produces the 't Hooft flux sector on the commutative torus
with $\Tr f = 2\pi$ and $\Tr f^2 = (2\pi)^2/N$.

Together they imply
\eqn\bpsfluxes{\eqalign{
\int\Tr f &\equiv 2\pi m = 2\pi q \sgn(p-q\theta)\cr
\int\Tr f^2 &= {(2\pi q)^2 V\over |p-q\theta|}.
}}

Non-zero electric flux $E_i$ and
internal momentum $P'_i$ will also contribute to the energy.
Their leading contribution is determined entirely by
the quadratic terms in \uoneact\ and the only effect on these
of turning on $\theta$ is a change in the overall normalization of the
action (from $\int \Tr 1$) leading to the same overall rescaling of their
contributions to the energy.  At least on the classical level, it is
easy to write explicit solutions with non-zero $E_i$
($A_i=E_i t$) or $P'_i$ (a plane wave with transverse polarization)
for which this is exact; these are BPS states in the gauge theory
and so this classical result should be exact.

This leads to the idea that both $P^i$ and $(P')^i$ are fixed under
variations of $\theta$.  This is only possible if $e_i$ varies and
the combination which remains fixed is
\eqn\modmom{
E_i \equiv e_i - \theta_{ij} P^j .
}
Such an effect is possible because the original action \action\
was a function only of the combination $X_i + A_i$.
Distinguishing the shift of the constant mode of $A_i$
generated by $E_i$,
from the shift of $X_i$ generated by $P^j$ requires making an choice
of convention, which could be $\theta$-dependent.

Adding these contributions and allowing
the topological terms \topterm\ and $\alpha \Tr 1$
leads to the final formula for the energy,
\eqn\bpsenergy{
E = {1\over |p-q\theta|} \left( g^{ij} E_i E_j
+ {V\over g_{YM}^2}(2\pi q)^2 \right)
+ \sqrt{g_{ij} (P')^i (P')^j} + \gamma m + \alpha |p-q\theta|.
}

In section 6, we will see that this result 
is symmetric under the $SL(2,\Zb)\times SL(2,\Zb)$ action
described in section 4.

\newsec{Physical interpretation}

We will shortly
propose an interpretation for the BFSS model compactified
on $\Tb^\theta$, generalizing that for the commutative torus $\theta=0$.
In this case the accepted interpretation \susslc\ is
that it is a non-perturbative
definition of M theory compactified on the manifold
$T^d\times (S^1\times\Rb)^{1,1} \times \Rb^{9-d}$
(we focus on $d=2$, but formulas in which $d$ appears are more general)
so we start with a short review of this theory.
(See \townsend\ for a review of M theory
covering the features we will use here.)

For many purposes, and in particular for understanding the classification of
topological sectors of this theory, we can think of M theory in terms of
its low energy limit, eleven-dimensional supergravity.
Eleven dimensional supergravity has
as bosonic degrees of freedom an eleven-dimensional metric $g_{AB}$ with
curvature scalar $R$,
and a three-form gauge potential $C_{ABC}$ with derived field strength
$G\equiv dC=4\partial_{[A}C_{BCD]}$ (indices $ABC\ldots$ are tangent
space indices).
The action is
\eqn\elevensg{
L_{11} = {1\over 2\kappa^2}\int d^{11}x\ \sqrt{g}\bigg(
-R + {1\over 2} G\wedge *G
+ {1\over 6} C\wedge G\wedge G
+ {\rm fermion\ terms.}
\bigg)
}
Besides general covariance the theory enjoys
local supersymmetry, acting as
\eqn\elevensusyall{\eqalign{
\delta e_I^\mu &= \half \bar\eta \Gamma^\mu \psi_I\cr
\delta C_{IJK} &= -{{3 \over 2}}
	\bar\eta \Gamma_{[IJ} \psi_{K]}\cr
\delta \psi_I &= D_I\eta + {1\over 2\cdot 12^2}
	G_{JKLM}(\Gamma_I^{JKLM}-8\delta_I^J\Gamma^{KLM})\eta  + fermi,
}}
where $\eta$ is a 32 component Majorana spinor,
$e^\mu_I$ an elfbein,
$\Gamma_A = \Gamma_\mu e^\mu_A$ are the Dirac matrices and
$\Gamma_{{I_1}\ldots {I_n}} \equiv
(1/n!)\sum_{\sigma\in S_n} (-1)^\sigma
	\Gamma_{I_{\sigma(1)}}\ldots \Gamma_{I_{\sigma(n)}}$
is an antisymmetrized product with weight one.
There is also a symmetry under gauge transformations
$\delta C = d\lambda$.

In direct analogy to the discussion of global symmetry in general
relativity (diffeomorphisms which preserve the background metric correspond
to Killing vectors), local supersymmetry transformations which preserve
the background are interpreted as global supersymmetries.
The usual case of interest is $\psi_I=0$ and in this case
supersymmetric vacua are characterized by the existence of solutions $\eta$
to
\eqn\elevensusy{
0 = \delta \psi_I = D_I\eta + {1\over 2\cdot 12^2}G_{JKLM}
	(\Gamma_I^{JKLM}-8\delta_I^J\Gamma^{KLM})\eta
}

Maximal supersymmetry is the case in which any constant $\eta$ is a solution
to \elevensg, which will be true if the Riemann curvature $R_{ABCD}=0$ and 
$G=0$.
The only such spaces are $T^d\times M^{1,10-d}$ with $T^d$ a torus
and $M^{1,10-d}$ a Minkowskian space (to start with, $\Rb^{1,10-d}$,
but we will modify this slightly below).\footnote*{
There is one other type of solution with maximal supersymmetry,
$S^d \times AdS^{1,10-d}$, but it will not be relevant in the present work.}

Thus the data (or `moduli')
of such a compactification are a flat metric on $T^d$,
and for $d>2$ a three-form tensor $C_{ijk}$ on $T^d$.
We will work with coordinates $x^i \cong x^i + 1$ for $1\le i\le d$
on $T^d$ and explicit components $g_{ij}$ and $C_{ijk}$.

For $d=2$ it is convenient to use instead the complex modulus $\tau$ and
volume $V$, for which the metric is
$$ds^2 = V |dx^1 + \tau dx^2|^2 .$$
The moduli space of compactifications is then $F\times \Rb^+$ where
$F \equiv SL(2,\Zb)\backslash SL(2,\Rb)/SO(2)$ is the usual fundamental domain.

For $d>2$, the analogous moduli space would be
$SL(d,\Zb)\backslash SL(d,\Rb)/SO(d) \times \Rb^+$.
However this is only a subspace of the moduli space, because of the additional
$d(d-1)(d-2)/6$ parameters $C_{ijk}$.  There will also
be additional identifications leading to the physical moduli spaces,
as we review in the next section.

The BFSS model is supposed to reproduce M theory but with the modification
$M^{1,10-d} \cong (S^1\times\Rb)^{1,1} \times \Rb^{9-d}$, a quotient of
$\Rb^{1,10-d}$ by a translation symmetry along a distinguished null
vector.
This space admits only a subgroup of $11-d$-dimensional Lorentz invariance,
$SO(1,1)\times SO(9-d)$, and there are additional moduli
compatible with this symmetry.  Our ultimate goal will be to explain
what these are and why compactification on the noncommutative torus
corresponds to turning on these moduli.

\subsec{BPS states and U-duality}

We now ask how the physics looks at length scales much larger than any
scale associated with the compactification $T^d$.  Such an observer will
see an effective space-time $M_{1,10-d}$
and dynamics governed by an action which (to zeroth approximation) is
obtained by restricting all fields in \elevensg\ to be constant on $T^d$.
The resulting theory will contain fields which transform as one-form
gauge potentials on $M_{1,10-d}$; clearly these will include
$g_{i\mu}$ and $C_{ij\mu}$.
These couple to particles carrying conserved charges which we
denote $e_i$ and $m_{ij}$ respectively.

Standard considerations show that $e_i$ is the usual
conserved momentum $e_i = (-i/2\pi)\p/\p x^i$, in our conventions
integrally quantized, and thus charged particles exist.  They are
simply particles of the quantized theory \elevensg\ with non-zero
internal momentum (or ``Kaluza-Klein modes'').

The particles carrying $m_{ij}$ are perhaps less familiar but this is
where the characteristic features of M theory start to appear.  The
Lagrangian \elevensg\ admits a wide variety of solitonic solutions
characterized by the charges $\int *G$ and $\int G$.
If $G$ were the two-form field strength of
four-dimensional abelian gauge theory,
these integrals would be electric and magnetic charges, respectively.
Although here $G$ is a fourth rank tensor, they share most of the
same properties:
non-trivial solutions must contain singularities of $C$,
but away from these singularities the charges are conserved and
satisfy a relative Dirac quantization condition.\footnote*{
Strictly speaking, the Chern-Simons term in \elevensg\ modifies 
the conserved `electric' charge to $\int *G + \half C\wedge G$ \page.}

The charges are defined as integrals over a seven-cycle and four-cycle
respectively and so the natural singularities are a $2+1$-dimensional
hypersurface and a $5+1$-dimensional hypersurface (respectively).
Such solutions are referred to as two-branes (more usually, membranes)
and five-branes; the definition of M theory includes the statement that
these two solutions (each carrying a quantized unit of charge)
describe well-defined objects in the theory.

Given this assumption, it follows that M theory
compactified on $T^2$ contains a particle with unit $m_{12}$ charge.
It is simply a membrane with the $2+1$ hypersurface taken to be
$T^2 \times \Rb$ for some time-like geodesic $\Rb \subset M^{1,10-d}$.
This is referred to as a ``wrapped membrane.''
Similarly, compactification on $T^d$ will contain particles with
any specified $m_{ij}=1$.
Wrapped five-branes will also correspond to particles for $d\ge 5$.

Both Kaluza-Klein modes and wrapped membranes are
BPS states -- although \elevensusy\ admits no solutions in the generic
configuration, for these configurations it does.
This is a far-reaching statement, some of whose implications we will
use, but we will not use its supergravity version in detail and
refer to \townsend\ for a complete discussion.
What we will use is the matrix theory version -- already described in
section 2 -- as well as the following implication: the energy of a BPS
state (in these theories) is exactly equal to the value computed classically.
Thus the energy of a Kaluza-Klein mode is determined by the usual
relation for a massless relativistic particle,
\eqn\bpskk{
E = |p| = \sqrt{e_i e_j g^{ij}} .
}

An equally explicit computation for the membrane would of course require
introducing the solution, but the result has a simple intuitive statement:
the energy of a wrapped membrane is equal to a constant membrane tension
multiplied by the area of the two-surface over which it is wrapped:
\eqn\bpsmem{
E = \sqrt{m^{ij} m^{kl} g_{ik} g_{jl} }.
}

We described in words the BPS states with unit charge, but BPS states
with general quantized charge can also exist and it is a dynamical question
whether or not they do.  However there is a very strong hypothesis which
leads to constraints: that of ``U-duality.''  Let us explain this in the
first non-trivial case of $d=3$.

{}From what we have said so far,
the moduli space of compactifications on $T^3$ should be
$$SL(3,\Zb)\backslash SL(3,\Rb)/SO(3) \times \Rb^+ \times \Rb, $$
with the last $\Rb$ factor corresponding to $\int C_{123}$.
The energy of a BPS state would be given by the sum of \bpskk\ and
\bpsmem.  As a physical observable this must be $SL(3,\Zb)$ invariant
and indeed both expressions together with the lattice of allowed charges
$(e_i,m^{jk})\in \Zb^6$ have manifest $SL(3,\Zb)$ symmetry.

However, the full analysis leads to three corrections to the previous
paragraph.  First, the two contributions \bpskk\ and
\bpsmem\ to the energy actually add in quadrature.
This is not hard to understand, by a standard argument relating the energy to
central charges of the supersymmetry algebra, but would require a detour and
we instead refer to \townsend.
Second, the complete energy formula has additional terms.  It
is\footnote*{for BPS states preserving $16$ supersymmetries;
in general there are further corrections.}
\eqn\bpstotal{
E^2 = (e_i + C_{iab} m^{ab}) (e_j + C_{jcd} m^{cd}) g^{ij}
 + m^{ij} m^{kl} g_{ik} g_{jl} .
}
The $Cem$ cross term comes from
a higher dimensional version of a familiar effect in standard
gauge theory.  In a constant gauge field background $A_i$,
the canonical momentum $p_i$ for a particle with charge $q$ is modified from
$(-i/2\pi)\p_i$ to $(-i/2\pi)\p_i + q A_i$.
Exactly the same happens here, with $q A_i$ identified with
$\int C_{iab}$.

The expression \bpstotal\
still has $SL(3,\Zb)$ symmetry, and a new $\Zb$ symmetry
$$\eqalign{
C_{123} &\rightarrow C_{123} + 1 \cr
e_i &\rightarrow e_i - \epsilon_{ijk} m^{jk} \cr
m^{jk}&\rightarrow  m^{jk},
}$$
directly analogous to those which lead to compactness of moduli spaces
of flat connections.
It is a particular case of the general statement that for compactification
on $\Mc$,
\eqn\genthreecom{
C \cong C' \ {\rm iff}\
\int_{\Sigma} C- C' \in \Zb
}
for every three-cycle $\Sigma\in \Mc$.\footnote{$^\dagger$}{
Physically, this is usually justified by observing that the action
of a `membrane instanton' wrapped on $\Sigma$,
a three-dimensional solution of the Euclidean form of \elevensg, will differ
by $2\pi i n$ between these configurations.
A mathematical discussion is given in \freed, where relations like
\genthreecom\ are 
made precise by interpreting equivalence classes of these objects
as elements in a smooth Deligne cohomology group.
See also \caicedo.
}

Finally,
by writing $m_i = \half\epsilon_{ijk} m^{jk}$, $V=(\det g)^{1/2}$ and
\bpstotal\ as
\eqn\bpstotaltwo{\eqalign{
E^2 &= (e_i + C_{iab} m^{ab}) (e_j + C_{jcd} m^{cd}) g^{ij}
 + V^2 m_i m_j g^{ij} \cr
&= (e_i + (C_{123}+iV) m_i ) (e_j + (C_{123}-iV) m_j) g^{ij} ,
}}
we see that it also has an $\Zb_2$ symmetry which acts as
$$\eqalign{
i V + C_{123} &\rightarrow -1/(i V + C_{123}) \cr
(\det g)^{-1/3} g^{ij} &\rightarrow
	(V^2 + C_{123}^2)(\det g)^{-1/3} g^{ij}  \cr
e_i &\leftrightarrow m_i .
}$$
This combines with the $\Zb$ symmetry to generate the group $SL(2,\Zb)$.

The complete symmetry group
\eqn\udualthree{
SL(3,\Zb) \times SL(2,\Zb)
}
is the U-duality group in $d=3$,
the largest discrete symmetry preserving the charge lattice and BPS
energy formula, and the non-trivial claim is that the full M theory
respects this symmetry, so that the true moduli space of compactifications
on $T^3$ is
$SL(3,\Zb)\backslash SL(3,\Rb)/SO(3) \times F$.
In particular, this implies that the multiplicity of BPS states for
each charge is invariant under U duality.
Much evidence has been found for this conjecture, and its many generalizations
to arbitrary $d$ and non-toroidal compactifications.

There are two ``proofs'' in the case of $d=3$.
The original argument came from the relation of this
theory to superstring theory.  The basic relation is that compactification
of M theory on $S^1$ produces the \IIa\ superstring theory, with the
membrane wrapped on $S^1$ becoming the string.  One
can show to all orders in string perturbation theory that this theory
enjoys a T-duality symmetry which acts on the BPS states as above.

Another argument comes from matrix theory, to which we turn.

\subsec{M theory in the IMF and Matrix theory}

The BFSS model defines M theory in the infinite momentum frame (IMF).
This means that only a subgroup $SO(1,1)\times SO(9)\subset SO(1,10)$
of Lorentz invariance is manifest.  Let $x^+$ and $x^-$ be two coordinates
on which $SO(1,1)$ acts by rescaling
$x^\pm\rightarrow \lambda^{\pm 1}x^\pm$ (so, $\p/\p x^\pm$ are null
Killing vectors).  Let $p_+=-i\p/\p x^+$ and $p_-=-i\p/\p x^-$ be the
conjugate momenta, so the usual relativistic relation $p^2 = m^2$ becomes
\eqn\lcpart{2 p_+ p_- = \sum_i p_i^2 + m^2 .}
The gauge theory Hamiltonian is interpreted as generating translation in $x^+$
(``light-cone time''), so energy in the sense of gauge theory becomes $p_+$.
The momentum $p_-$ is then identified with $N$, the rank of the gauge
group, as $p_-=N/R$ where $R$ is the normalization parameter in \bfssact,
sometimes called the ``radius of the light-cone dimension $x^-$.''

We first explain the description of the BPS states we described
above.  Particles in the supergravity multiplet are massless and
\lcpart\ becomes $p_+ = R/2N \sum_i p_i^2$.
For $N=1$ this is the dynamics of a single eigenvalue of the matrix
governed by the quantum mechanics \bfssact.
To get the entire spectrum,
there must be a unique zero energy bound state in matrix
quantum mechanics for each $N>1$.  The ``center of mass'' degrees
of freedom $\Tr X$ have free dynamics and a state with center of
mass momentum $p_i=\Tr P_i$, using $P_i = {\bf 1} p_i/N$, will
have $p_+ = R/2 \sum \Tr P_i^2 = R/2N \sum_i p_i^2$.

More generally, we could consider models with the same spectrum
of zero energy  bound states, but in which the center of mass
is described by supersymmetric quantum mechanics on a target space $\Mc$.
These will have an interpretation as M theory compactified on
$\Mc \times (S^1\times\Rb)$.

The additional charges $e_i$ and $m^{ij}$ of toroidal compactification
must be identified with the additional conserved charges of $p+1$
super Yang-Mills theory.
The correspondance is
\eqn\matrixcharges{\eqalign{
N		\qquad& p_- \cr
\int F_{0i}	\qquad&e_i \qquad {\rm electric\ charge}\cr
\int F_{ij}	\qquad&m_{ij} \qquad {\rm magnetic\ charge} \cr
}}

We can now state the matrix theory argument
\refs{\grt,\suss}\ for the U-duality group \udualthree.
{}From the general discussion
in section 3, the BFSS model compactified on $T^3$ is a $U(N)$ super
Yang-Mills theory in $3+1$ dimensions with maximal supersymmetry,
compactified on the dual torus $\tilde T^3$.
The $SL(3,\Zb)$ acts in the obvious way on $\tilde T^3$.
Furthermore, this $3+1$ SYM
is believed to enjoy an $SL(2,\Zb)$ duality symmetry,
acting on the charged states and parameters precisely as above.
The combination $iV+C_{123}$ is identified with the complex gauge
coupling.

\subsec{Duality in M theory with a compact null dimension}

How does the discussion of section 6.1 change if we take into
account all the moduli which preserve the IMF subgroup of Lorentz
symmetry $SO(1,1)\times SO(9-d)$ ?

The maximally supersymmetric backgrounds will again be characterized
by constant $G_{AB}$ and $C_{ABC}$, but now we can allow $A=+$ or $A=-$
in addition to the previous $1\le A\le d$.

The deformations with $A=+$ are physically trivial as this
dimension is not compact.  By suitable choice of coordinates and gauge
transformation we can set them to zero.  On the other hand,
it is useful to keep them with this understanding.

The deformations $g_{-i}$ and $C_{-ij}$ are non-trivial.
Turning on $g_{-i}$ would be very interesting but we will confine ourselves
to a short comment about this at the end of the section.

What we will claim is that turning on $C_{-ij}$ corresponds
to deforming the commuting torus to the noncommuting torus.
More precisely, we will make the following identification:
\eqn\guess{
R \int dx^i dx^j C_{ij-} = \theta_{ij} .
}
The constant of proportionality in this relation is determined by
identifying the periodicity $\theta_{ij} \sim \theta_{ij}+1$
with the periodicity \genthreecom\ and $R = \int dx^- 1$.

In this and the next subsection
we discuss M theory properties of these additional moduli.
In particular, we ask whether the U-duality group for compactification on $T^2$
is larger than $SL(2,\Zb)$.
Clearly it will be a subgroup of that for $T^3$ and
the maximal subgroup which would preserve the distinguished
direction $x^-$ is the subgroup $SL(2,\Zb)_C \subset SL(3,\Zb)$ times
the non-classical $SL(2,\Zb)_N$.\footnote{$^1$}{
More precisely, the full duality group should include additional
inhomogeneous transformations to become a discrete
subgroup of a contraction of $SL(3,\Rb)$ \hull.}

We conjecture that a non-classical $SL(2,\Zb)_N$, generated by
the transformations $\theta\rightarrow\theta+1$ and
$\theta \rightarrow -1/\theta$, is present.
Our original motivation for this claim was the relation to gauge
theory on the noncommutative torus and the relation \guess.
In section 4 we saw that the Teichmuller space for the noncommutative
torus admitted two commuting $SL(2,\Zb)$ actions, which will become
exactly the $SL(2,\Zb)_C$ and $SL(2,\Zb)_N$ actions in the matrix
theory interpretation.

Let us go on however to discuss arguments purely in the context of M
theory, before returning to this interpretation.
Now there is already evidence that the multiplicities of BPS states
can have such enhanced duality symmetries \hack.
Indeed, we will propose an $SL(2,\Zb)_N$ S-duality action on
the charges which is a
particular case of the U-duality proposed there (reduced from $T^3$
to $T^2$ compactification).  We will be able to go on and propose an
action on the moduli space which leads to a symmetry of the mass formula
and thus is a candidate for an exact duality of M theory, but only
in the case $\theta\ne 0$.

It is pointed out in \hack\ that the action of
the full U-duality group involves an additional class of BPS state --
membranes wrapped about one transverse
dimension (say $x^i$) and the $x^-$ dimension, or longitudinal membranes.
These also correspond to particles which carry a new conserved charge;
let us call it $m^{i-}$.  For zero $C_{ij-}$ their contribution to the
mass formula is known (following \hack) and the simplest possibility is that
it is
independent of $C_{ij-}$, which is compatible with the duality.\footnote{$^2$}{
This leads to a mass formula slightly
different from that given in the original version of this work.
We thank P.-M. Ho for a question on this point.}
We have not derived this
independently from M theory, but it would follow from the
noncommutative gauge theory under the assumptions in section 5.

The mass formula for BPS states then becomes
\eqn\lcdisp{\eqalign{
2 (p_+ + \Omega)p_- =  &p_\perp^2
+ {1\over V\tau_2}\left|\tau E_1 + E_2\right|^2
+ V^2 m^2 \cr
&+ \sqrt{{V\over\tau}} \left| w^{1} + \tau w^{2} \right|
}}
with $p_- = n/R - C_{-ij} m^{ij}$, $E_i = e_i - R C_{-ij} m^{j-}$
and $w^i = n m^{i-} - m^{ij} e_j$.
The term $\Omega$ stands for an arbitrary linear term
$$\Omega = \alpha p_- + C_{+12} m + g_{+i} e_i + C_{+-i} m^{i-}$$
which can be produced by modifications to the Hamiltonian such as \topterm:
$$H = H_0 + \int \Tr (\alpha + C_{+12} F_{12} + g_{+i} F_{0i}
+ C_{+-i} P^{i}).
$$
The term $\alpha$ corresponds to an additional boost in the
light-cone plane, while the other terms are
trivial background fields, which could be eliminated by gauge transformations.

We now apply the transformation
\eqn\sduality{\eqalign{
&\theta \rightarrow -1/\theta; \qquad
V\rightarrow V \theta^{2a}; \qquad R \rightarrow R \theta^b\cr
&\left(\matrix{n& e_1& e_2\cr m& m^{2-}& m^{1-}}\right) \rightarrow
\left(\matrix{0& -1\cr 1& 0}\right)
\left(\matrix{n& e_1& e_2\cr m& m^{2-}& m^{1-}}\right)
}}
From this follow
$$p_- \rightarrow \theta^{-1-b} p_-; \qquad
E_i \rightarrow E_i/\theta ; \qquad
w^i \rightarrow w^i .
$$
Then \lcdisp\ becomes
\eqn\lcdispdual{\eqalign{
2\left( p_+ + \tilde \Omega\right)
	p_- =  &\theta^{1+b} p_\perp^2
+ \theta^{-1-2a+b}{1\over V\tau_2}\left|\tau E_1 + E_2\right|^2
+ \theta^{1+4a+b}{V^2} n^2 \cr
&+ \theta^{1+a+b} \sqrt{{V\over\tau}}
	\left| w^1 + \tau w^2 \right|.
}}
where $\Omega$ transforms into $\tilde\Omega$ in an obvious way.

The $n^2$ term can be turned into a $\theta^2 m^2$ term
like the one appearing in \lcdisp\ by
adding a term $\tilde\Omega \propto (n+\theta m)$.
The powers of $\theta$ will then cancel if $a=-2/3$ and $b=-1/3$,
and we rescale $p_\perp\rightarrow \theta^{-1/3} p_\perp$.
A simple relation following from \sduality\ is
$R V^2 \rightarrow R V^2/\theta^3$.

Thus this combined transformation of the parameters is a symmetry of
the BPS spectrum.
As we discussed above, although the terms in $\tilde\Omega$ do have
interpretations as backgrounds, they are not really gauge invariant.
The gauge invariant physical predictions are expressed in \sduality.

\subsec{Relation to T-duality}

Another argument for duality under $SL(2,\Zb)_N$ uses the relation
to string theory.  The basic relation is that M theory compactified on
$S^1\times \Mc$ becomes type \IIa\ superstring theory on $\Mc$.
By analogy with the $p=3$ case we can try to interpret the S-duality
transformation \sduality\ as double T duality, but now acting on the
null torus $S^1 \times S^1_-$.

This fits well with the proposed action $\theta\rightarrow -1/\theta$.
We want to interpret it as the zero volume limit of the usual T-duality
relation $iV+B \rightarrow -1/(iV+B)$, and indeed the volume of a null
torus is $V=0$.

There is a strong analogy to the case of T-duality with a
compactified time dimension, discussed by Moore.
In  \moore\ it was shown that in this case
T-duality acts ergodically on the moduli space, and the
relevance of noncommutative geometry to this situation
was even pointed out!

Let us proceed to verify the T-duality by world-sheet computation.
The bosonic part of the action for a type \II\ string on
$S^1_b\times S^1_-\times \Rb$ will be
\eqn\stringact{
S =
\int {2R\over\alpha'} \p X^+ \pb X^-
+ {R_b^2\over\alpha'} \p X^b\pb X^b + \theta \p X^-\pb X^b
}
(where $\alpha'={l_p^3\over R_a}$ and we take $0\le X^-,X^b\le 2\pi$).
We then T-dualize the coordinates $(X^-,X^b)$ in the usual way,
which results in an action written in terms of the inverse metric
$$(G+B)^{-1} =
\left(\matrix{0&\theta\cr -\theta&{R_b^2\over\alpha'}}\right)^{-1}
= {1\over \theta^2}
\left(\matrix{{R_b^2\over\alpha'}&-\theta\cr \theta&0}\right) .
$$
As noted before, this transformation is quite singular for $\theta=0$.

We now take $\tilde X^-$ and $\tilde X^b$ to be the new null and space-like
coordinates (implicitly exchanging the two indices)
and find the action
$$
S =
\int {2\tilde R\over\alpha'} \p X^+ \pb \tilde X^-
+ {2\theta\tilde R \tilde R_b^2\over\alpha' } \p X^+ \pb \tilde X^b
+ {\tilde R_b^2\over \alpha'} \p \tilde X^b \pb \tilde X^b
- {1\over \theta} \p \tilde X^-\pb \tilde X^b
$$
with $\tilde R = R/\theta$ and $\tilde R_b = R_b/\theta$.

The end result is the expected effect $\theta\rightarrow -1/\theta$.
To get an M theory relation for the old and new radii,
let us combine the two transformation laws we found, keeping in mind
that the combinations which appear in the action (and should transform
at fixed $\alpha'$) are $R_b^2/\alpha'$ and $R/\alpha'$.
This suggests the combined transformation
$${R V^2\over l_p^6} =
\left({R R_a\over l_p^3}\right) \left({R_b^2 R_a\over l_p^3}\right)
\rightarrow 
\left({R R_a\over \theta l_p^3}\right) 
\left({R_b^2 R_a\over \theta^2 l_p^3}\right)
= {R V^2\over \theta^3 l_p^6}.
$$
This agrees with the result we found in the previous subsection.
In particular it is symmetric in $R_a$ and $R_b$, a non-trivial test.

\subsec{Matrix theory on the noncommutative torus}

To summarize sections 3 and 4, we found that we can deform the
commutative torus $T^d$ to a noncommutative
torus specified by the metric $g_{ij}$ and $d(d-1)/2$ additional parameters
$\lambda_{ij} = \exp 2\pi i \theta_{ij}$.
We now propose to define a matrix theory in terms of the corresponding
gauge theories of section 5,
in exactly the way conventional gauge theory is used.
In particular we take the gauge coupling $g_{YM}^2 = 1/V$,
and interpret the parameters $(p,q)$ (in $d=2$)
of the module $\Hc^{p,q}$
as conserved charges in space-time.

Since these theories are continuous deformations
of the theory on the commuting torus, it is
plausible that the same spectrum of zero-brane bound states exists
and that the space-time interpretation is a deformation of that
for the commuting torus.
The center of mass degrees of freedom are the transverse $\Tr X^\mu$
and the choice of constant curvature connection.
As we discussed, the moduli space of constant curvature connections
is a commuting torus with flat metric,
and this commuting torus is the target space.
However, since these theories have different physics from the
standard toroidal compactifications,
they must correspond to compactification on tori with background fields.
The existence of BPS states preserving $16$ supersymmetries
requires $G=0$, so the background fields can only be a constant
three-form tensor $C$.

The background $C_{ij+} \ne 0, C_{ij-}=0$ has an evident
realization in matrix theory.  By the usual rules of canonical quantization,
the LC Hamiltonian will be $P_+ + C_{ij+} m^{ij}$,
and such a term can be added to the Hamiltonian directly --
it corresponds to the topological term \topterm\ in the action.

Thus we conjecture that $\theta_{ij}$ corresponds to the background
$C_{ij-}$ as in \guess.
The natural generalization of \matrixcharges\ is to identify
\eqn\ncmatrixcharges{\eqalign{
N\dim \CH_{p,q}^\theta		\qquad& p_- \cr
\int F_{0i}	\qquad&e_i \cr
\int F_{ij}	\qquad&m^{ij}  \cr
\int T_{0i}	\qquad&m^{i-}  .
}}
In particular, we identify the parameters of the module $\Hc_{p,q}$
as $m=Nq=\Tr F_{12}$, and $n=Np$ as the canonical momentum for $\theta=0$.
The dimension formula \nctwodim\ then becomes
\eqn\dimcharges{
p_- = {n\over R} - C_{-ij} m_{ij}
}
which corresponds precisely to the expected contribution of a
state with membrane number $m_{ij}$ to $p_-$.

The formula \bpsenergy\ for the energies of BPS states in the
gauge theory,
then precisely reproduces \lcdisp, the expected energies in M theory,
and we have already verified the
$SL(2,\Zb)_C \times SL(2,\Zb)_N$ symmetry of this formula under \sduality.
Thus, if gauge theory on the noncommuting torus can be quantized
respecting these symmetries (and satisfying our other assumptions),
the U-duality of M theory on a null circle will follow from this
matrix theory definition.

Although we made certain assumptions in section 5, there are several
unambiguous predictions of the formalism (most notably, the formula
\dimcharges)
which already serve as non-trivial tests of the conjecture.

It can be shown that classically, the volume of the moduli space of flat
connections transforms as $V \rightarrow V/\theta^2$.  This
fits with the scaling in \sduality\ in the following sense.  
The rescaling of $p_\perp$ implies
that we are rescaling all transverse lengths by $l\rightarrow \theta^{1/3} l$;
if we follow that with $V \rightarrow V/\theta^2$ we get \sduality.

Let us mention another test which may be possible with the classical
theory.
This is to repeat the discussion of \dhn\ of light-cone gauge fixing for
the supermembrane in this background, and see if the result is equivalent to
\uoneact.  The problem is
that the three-form coupling $\int C_{-ij} dX^-\wedge dX^i \wedge dX^j$
involves the coordinate $dX^-$, which is determined by differential
constraints, leading to an apparently more complicated non-local action.
It would be quite interesting to prove or disprove the equivalence of these
two actions.

Finally, we note that the $V\rightarrow 0$ limit of the $2+1$-dimensional 
theory should be interpreted as \IIb\ superstring theory \Banks, and the
parameter $\theta$ will become a mixed component $g_{10,-}$ of the metric.
In a sense, this interpretation exchanges the roles of the two commutative
tori of section 4; $SL(2,\Zb)_C$ becomes a non-classical duality, while
the even torus becomes the target space.
The conjectured full Lorentz invariance of this model would follow if
in this limit (large dual volume and strong coupling), 
the low energy physics becomes independent of $\theta$. 

\newsec{Conclusions}

In this work we have described a specific connection between M
theory and noncommutative geometry. 

From the matrix theory point of view, we showed that
the noncommutative torus appears naturally, 
on the same footing as the standard
torus, to yield new solutions to the problem of toroidal
compactification of the BFSS model. 

We then gave a concrete description of this generalization,
in terms of a class of deformations of gauge theory characterized by
an additional two-form parameter.  The existing determination of the
Teichmuller space of flat noncommutative tori 
(in dimension 2) admits a natural
action of $SL(2,\Zb) \times SL(2,\Zb)$,
which suggests a corresponding duality in the associated gauge theories. 
Given certain plausible assumptions, the masses of BPS states in these
theories indeed have this duality symmetry.

Finally, we argued that M theory
compactification on a torus and a light-like circle
has a very similar generalization,
which had not been considered previously.
It is also determined
by a two-form parameter -- the integral of the three-form
of M theory along the light-like circle.
We found evidence for an $SL(2,\Zb) \times SL(2,\Zb)$
duality symmetry in these compactifications, both
in that it is a symmetry of the mass formula for certain BPS states,
and in that it is a sensible form of
T-duality in the related \IIa\ superstring theory.
Since this is an allowed background,
in the context of matrix theory the question
``what deformation of gauge theory corresponds to this generalization''
deserves an answer.
The similarity of the two generalizations 
lead us to the conjecture that gauge theory on the noncommutative torus
is the answer.  

In the absence of other candidates,
a true believer in matrix theory might even regard this as significant
evidence for its existence as a quantum theory.
However, we should not rest satisfied with this argument, as 
these theories have a quite concrete definition \uoneact\ and
it is fairly clear how to decide whether or not they are perturbatively
renormalizable.
We have only made a preliminary investigation of this question
and can only state that the obvious arguments against it (for example,
that these are higher derivative theories with a priori bad ultraviolet
behavior) appear to be simplistic.

In addition to the matrix theory motivation
it would clearly be quite important to find any
sensible deformation of maximally supersymmetric gauge theory.
These theories in a sense allow continuously
varying the rank $N$ of the gauge group, 
and realize symmetries relating sectors of different $N$.
Furthermore, since they are particularly simple (non-'t Hooft)
large $N$ limits of conventional
gauge theory, they could also be interesting in the study of
the large $N$ limit in general.

This very specific link between M theory and noncommutative geometry suggests
that noncommutative geometry could be the geometrical framework in which 
M theory should be described.  
One way to carry this forward would be to translate matrix theory into
the framework of spectral triples in noncommutative geometry.
This is based on the Dirac operator $D \equiv \Gamma_i X^i $ and a
simple equation characterises the $D$'s corresponding to commutative and
noncommutative spaces.

The physical test of the framework will be to see if the natural constructions
it suggests have sensible physical interpretations.
A straightforward generalization of the work here would be to 
use deformations of gauge theory on a curved manifold parameterized by
a Poisson structure, or closed two-form.  We expect that these will
have the same interpretation as closed three-form backgrounds in M theory.
It will be quite interesting to see if the
new features of matrix theory apparent in compactification on
$T^p$ with $p>3$ or on curved space have equally direct analogs in the
framework of noncommutative geometry.

\bigskip
\noindent
{\bf Acknowledgements}
\medskip
We thank Bernard de Wit,
Dan Freed, Pei-Ming Ho, Chris Hull, Marc Rieffel,
Samson Shatashvili and Matthias Staudacher
for useful discussions and correspondance.

\listrefs\end